\documentclass[9pt,twocolumn,twoside]{osajnl}

\journal{josab} 

\setboolean{shortarticle}{false}
\usepackage{braket}
\usepackage{blkarray}
\usepackage{xcolor}

\usepackage{soul}
\usepackage{ulem}
\def\bcen{\begin{center}}
\def\ecen{\end{center}}
\usepackage{epstopdf}
\ifthenelse{\boolean{shortarticle}}{\colorlet{color2}{color2b}}{\colorlet{color2}{color2}} 

\title{State transfer with separable optical beams and  variational quantum algorithms with classical light}

\author[1,*]{Sooryansh Asthana}
\author[1,$\dagger$]{ V. Ravishankar}

\affil[1]{Department of Physics, Indian Institute of Technology Delhi, New Delhi, 110016, India.}
\affil[*]{Corresponding author: sooryansh.asthana@physics.iitd.ac.in}
\affil[$\dagger$]{vravi@physics.iitd.ac.in}

\dates{Compiled \today}

\ociscodes{(000.1600) Classical and  quantum physics  (270.0270)  Coherence (200.3050)   Information processing (270.5585) Quantum information and processing }

\doi{\url{http://dx.doi.org/10.1364/ao.XX.XXXXXX}}

\begin{abstract}
Classical electromagnetic fields and quantum mechanics-- both obey the principle of superposition alike. This opens up many avenues for simulation of a large variety of phenomena and algorithms, which have hitherto been considered quantum mechanical.  In this paper, we propose two such applications. In the first, we introduce  a new class of beams, called  equivalent optical beams, in parallel with equivalent states introduced in [Bharath \& Ravishankar,  \href{https://doi.org/10.1103/PhysRevA.89.062110}{Phys. Rev. A 89, 062110}]. These beams have the same information content for all practical purposes. Employing them, we show how to transfer information from one degree of freedom of classical light to another, without any need for classically entangled beams. Next, we show  that quantum machine learning can be performed with OAM beams through the implementation of a quantum classifier circuit.  We provide explicit protocols and  explore the possibility of their experimental realisation. 
\end{abstract}

\setboolean{displaycopyright}{true}

\begin{document}

\maketitle
\thispagestyle{fancy}

\ifthenelse{\boolean{shortarticle}}{\ifthenelse{\boolean{singlecolumn}}{\abscontentformatted}{\abscontent}}{}
\section{Introduction}
 \label{Introduction}    
Recent times have witnessed a surge of interest in quantum algorithms and quantum foundations. This has culminated in quantum computing algorithms \cite{Deutsch85, Deutsch92, Simon97, Berstein97}, quantum factorisation algorithm \cite{Shor97}, quantum search algorithm \cite{Grover96} and variational quantum algorithms \cite{wan2017quantum, farhi2018classification, rebentrost2018quantum, schuld2019quantum, havlivcek2019supervised}. The reason for quantum speedups in these algorithms has been a matter of intense research so as to identify appropriate resource states. This, in turn, has led to significant developments in resource theories of quantum entanglement \cite{Horodecki09}, quantum coherence \cite{Streltsov17}, quantum discord \cite{Bera17}, quantum contextuality \cite{amaral2019resource}, etc. From the viewpoint of quantum foundations, it is of paramount importance to  differentiate the effects that are exclusively quantum mechanical vis-a-vis that can be simulated by some classical systems \cite{Ferrie14}.
 
 In parallel, there have been remarkable developments in generation, manipulation and detection of orbital angular momentum (OAM) modes of light \cite{Allen1999iv, andrews2012angular, allen2011orbital, marrucci2011spin}. These modes have found applications in diverse fields, e.g., optical tweezers, microscopy and imaging (see \cite{shen19} and references therein). Of particular importance to us are the following two facts: (I) a classical beam carrying OAM of light provides an experimental demonstration of a higher dimensional system obeying the principle of superposition, and, more significantly, (II) SU(2) coherent OAM beams have been experimentally generated in \cite{tuan2018realizing}. These beams are classical counterparts of $SU(2)$ coherent states in quantum mechanics. 
    
In classical computation, there is no difference between $N$ bits and a single dit with $d=2^N$, as far as algebraic manipulations are concerned. So, the representations in different number systems, e.g., binary, octal, decimal and hexadecimal systems can be freely converted to one another. The situation changes in quantum computation due to the principle of superposition. Additionally, in the quantum realm, two subsystems of entangled states may be at different locations. Barring that, classical waves and quantum systems share common features. 

A systematic study of those features, that are shared with classical systems, bears a twofold advantage. From the foundational perspective, it bifurcates the exclusively-quantum mechanical features and the quantum mechanical features shared by some classical systems as well. From the viewpoint of applications, this study provides us with various platforms for  implementation of many quantum algorithms and protocols, say, with OAM modes of classical light. 
 In fact, electromagnetic fields have different physical properties associated with them. Interference involving these degrees of freedom has been demonstrated in \cite{Mandel95, sztul2006double, Wolf07, jha2010angular, Malik12measurement}. For example, a two--dimensional Hilbert space in quantum mechanics shares a formal equivalence with a two-dimensional space spanned by the two orthogonal polarisations of the classical light. OAM modes of light \cite{Allen92} form a theoretically unbounded space with discrete indices. The coherent mode representation of partially coherent light fields is formally equivalent to the density matrix in 	quantum mechanics \cite{Wolf07}.

 Consider a classical beam described by, say, its polarisation, $|\hat{n})$  and momentum, $|\vec{k})$. If the beam admits an expansion $\int_k d\vec{k}|\vec{k}, \hat{n}_k)(\vec{k},\hat{n}_k|$, it is termed as {\it separable}, otherwise it is termed as {\it entangled} \cite{Forbes19}. For example, structured vector beams of light are inseparable in polarisation and OAM degrees of freedom. There have been many remarkable developments in   simulation of quantum phenomena with classically entangled beams. For example, violation of Bell--like inequality has been demonstrated with optical vortex (Laguerre-Gauss) beams in \cite{Stoklasa15} by using their expansion in terms of Hermite--Gauss modes. Similarly, a Bell--like inequality has been proposed and experimentally tested for spin--orbit inseparability of a laser beam \cite{Borges10}.  In \cite{Kagalwala13}, it has been experimentally verified that classical entanglement between two degrees of freedom is quantified by  Bell's measure.

 This suggests that those quantum  phenomena or algorithms, that have their geneses in parallelism and entanglement at one spatial location \cite{Forbes19}, can also be implemented with classical electromagnetic waves. 
It has led to an ongoing interest to implement quantum algorithms and protocols with orbital angular momentum modes of classical light. For example, in \cite{Hashemi15}, classical analogue of quantum teleportation has been demonstrated for transfer of information from OAM to polarisation degree of freedom of a beam.  In \cite{Li18},  monogamy relation between contextuality and nonlocality in quantum theory has been simulated in classical optical systems. It has been performed by using polarization, path, and OAM of a classical optical beam.  Quantum random walk has been implemented in \cite{Goyal13}, by employing OAM and polarisation degrees of freedom of classical light.  The proposed experimental setups accomplish the same task without the need for a single photon source, which is hard to prepare \cite{eisaman2011invited}. There are many more quantum phenomena and algorithms that have been simulated with classical light, but not necessarily with OAM modes \cite{Hemmer06, Kaur07, Dixon09, Atherton15, Perezgarcia16, Garcia18, Konrad19, Zhang19, Chevalier21}. For a detailed review of classical entanglement, one may refer to \cite{Forbes19} and references therein. We refer the reader to \cite{Konrad19} for a discussion of quantum mechanics and classical electromagnetic theory.

 In this paper, we take a step ahead and ask if there can be a further simplification. We start with the question: whether classically entangled beams can themselves be replaced by classically separable beams in information theoretic processes, perhaps, in higher dimensions? As mentioned earlier, by classically separable beams, we mean those beams that can be obtained as incoherent superpositions of pure separable beams. To answer this question, we first introduce equivalent beams, analogous to  equivalent states, proposed in \cite{Bharath14} in the quantum domain.  Equivalent beams are best understood through {\it $SU(2)$--coherent beams}. $SU(2)$ coherent beams share all the properties of $SU(2)$ coherent states. These beams have been experimentally realised recently with OAM modes \cite{tuan2018realizing}. The set of $SU(2)$- coherent beams forms an overcomplete basis. So, any beam can be expanded in this basis. The expansion of any beam in this basis is termed as its {\it Q-- representation}. Two optical beams are recognized to be {\it equivalent} if they share the same Q-- representation. Interestingly,  higher--dimensional (e.g., OAM) equivalent beams of a pure lower dimensional (e.g., fully polarised) beam turn out to be highly mixed. More importantly,  equivalent beams of a lower dimensional classically entangled beam can, in fact, be separable. Thus, this formalism provides us with a tool to harness highly mixed separable beams for information processing.
 
  As an application of this formalism, we propose a protocol for transferring information from the path degree of freedom to the OAM degree of freedom of a classical beam. This protocol is the classical optical version of a protocol, recently proposed by us in \cite{Asthana21} in the quantum realm.  Noting the advancements in generation, manipulation and detection of OAM modes \cite{Allen1999iv, andrews2012angular, allen2011orbital, marrucci2011spin}, we believe that this protocol may be implemented experimentally.

 Many variational quantum algorithms \cite{wan2017quantum, farhi2018classification, rebentrost2018quantum, schuld2019quantum, havlivcek2019supervised} also require generation, manipulation and detection of quantum states, all at the same spatial location. Hence, they may also be implemented with classical optical beams. As a concrete example, we outline an  experimental setup of the quantum classifier circuit, proposed in \cite{Adhikary20a}, by employing OAM modes of classical light. The experimental setup would require a  spatial light modulator \cite{Zhang2020review} and a programmable hologram \cite{Wang17}, both of which are available.

The plan of the paper is as follows: in section (\ref{Notation}), we setup the notation to be used throughout the paper.  For the sake of clarity and completion, the paper is written in somewhat pedagogical style.  In section (\ref{Formal_mapping}), the formal mapping between classical electromagnetic fields and quantum states is discussed. The central results of the paper are contained in sections (\ref{Equivalent}), (\ref{Remote state preparation unknown qubit}) and (\ref{meth}).  In section (\ref{Equivalent}), the concept of equivalent beams is proposed, followed by the protocol for transfer of information using separable equivalent of a  classically entangled beam in section (\ref{Remote state preparation unknown qubit}).   In section (\ref{Information_noise}), we show how  information transfer can be accomplished between different degrees of freedom of a noisy beam, by employing a detector with higher resolution for its retrieval. Sections (\ref{VQA}) and (\ref{meth}) discuss simulability of a variational quantum algorithm with OAM modes of classical light. 
Section (\ref{Conclusion}) summarises the paper with concluding remarks.

\section{Notation}
\label{Notation}
We first setup the notation for an uncluttered discussion.
\begin{enumerate}
\item  Following \cite{Spreeuw01}, we employ $|)$ and $(|$  for classical optical beams,  in place of ket and bra respectively, in order to avoid confusion.  We employ Dirac's Bra-Ket notation for quantum mechanical states.
\item We use the symbol $J$ for coherent mode representations of classical beams and reserve the symbol $\rho$ exclusively for quantum mechanical density matrices. The superscript of $J$ represents the degree of freedom which the coherent mode representation belongs to, which will be clear from the context.
\item The symbol $|{\rm LG}_{pl})$ is used for Laguerre-Gauss modes with the radial mode index $p$ and the azimuthal mode index $l$.
\item The generators of $SU(2)$ in any dimension will be represented by $T_1, T_2$ and $T_3$. In two dimensions,  the generators $T_i$ have the form  $T_i\equiv \frac{\sigma_i}{2}$, where $\sigma_i$ are Pauli matrices.
\item  We shall, henceforth, employ the following two shorthand notations, in order to avoid mathematical clutter:
\begin{align}
    \vec{T}\cdot\hat{m} &\equiv T_1m_1+T_2m_2+T_3m_3,\nonumber\\
    \sum_{i=1}^3T^A_i\otimes T^B_i&\equiv\vec{T}^A\cdot\vec{T}^B \equiv T^A_1T^B_1+T^A_2T^B_2+T^A_3T^B_3,
\end{align}
where $m_1, m_2, m_3$ are components of $\vec{m}$ along three mutually orthonormal directions $\hat{e}_1$, $\hat{e}_2$ and $\hat{e}_3$ respectively. In the quantum domain, the symbols $A$ and $B$ can represent two subsystems of the same system whereas for  classical light, they only represent two degrees of freedom of a light beam. For example, $A$ and $B$ may represent polarisation and OAM degrees of freedom respectively.
\end{enumerate}

\section{Formal equivalence between quantum mechanical states and classical electromagnetic fields}
\label{Formal_mapping}
We briefly recapitulate the polarisation matrix of a classical electromagnetic field, given in \cite{Mandel95, Wolf07}. 
We represent the coordinates of the electric field associated with a quasimonochromatic plane wave $(z = 0)$,  in terms of a row vector as,
\begin{align}
{\bf E} = E_x(t)\hat{x}+E_y(t)\hat{y} \mapsto \begin{pmatrix}
E_x(t), & E_y(t)
\end{pmatrix}.
\end{align}
 If the field is not completely coherent, we can represent it in terms of a $2\times 2$ density matrix $J^{\rm pol}$, termed as the {\it polarisation matrix}. In the $x-y$ coordinate system, it takes the following form,
\begin{align}
J^{\rm pol}= \dfrac{1}{I}\begin{pmatrix}
J_{xx} & J_{xy}\\ J_{yx} & J_{yy}
\end{pmatrix} = \dfrac{1}{I}\begin{pmatrix}
\langle E^*_x(t) E_x(t)\rangle & \langle E_x^*(t)E_y(t) \rangle \\ \langle E^*_y(t) E_x(t)\rangle & \langle E_y^*(t)E_y(t) \rangle 
\end{pmatrix}.
\end{align}
The angular bracket $\langle \cdot\rangle$ represents the time-average and $I$ represents the total intensity. Note that the polarisation matrix, $J^{\rm pol}$, is a positive-semidefinite hermitian matrix with unit trace. So, in its eigen-basis,  the matrix $J^{\rm pol}$ can be resolved as an incoherent sum of two coherent modes. If it represents a  fully coherent field, one of the eigenvalues vanishes.

\subsection{Formal analogy between polarisation matrix of classical light and a qubit: coherent mode representation}
\label{Coherent_mode_representation}
In classical electromagnetic fields, all the components of the fields are, in principle, completely measurable. Quantum mechanics, however, starts with probability amplitudes, whose bilinears are measurable. For clarity and pedagogical purpose, it is, therefore, worthwhile to establish what the formal equivalence between the two is. 

Fully and partially coherent electromagnetic fields are formally equivalent to pure states and mixed states in quantum mechanics respectively. It is, however, crucial to note that this mapping is purely formal in the sense that quantum mechanical probability amplitudes map to the amplitudes of electromagnetic fields. Quantum mechanical probabilities, intrinsic to quantum mechanics, map to normalised intensities of the  electromagnetic fields.  We, thus, recognise that $J^{\rm pol}$ has the same structure as that of the density matrix of a qubit, though the underlying physical quantities are completely different (for an illuminating discussion of partially polarised light, we refer the reader to \cite{landau2013classical}).

\subsubsection{c-Entropy as a quantifier of degree of polarisation}
 Since probability in quantum mechanics maps to normalised intensity of an electromagnetic field, all the functions of quantum probability naturally map to corresponding functions of normalised intensity. So, the classical counterpart of von--Neumann entropy $S(\rho) = -{\rm tr}(\rho{\rm log}\rho)$ of a qubit $\rho$, is given by a function, that we term as {\it $c$--entropy},
\begin{align}
{\cal S}(J^{\rm pol}) = -{\rm tr}(J^{\rm pol}{\rm log} J^{\rm pol}).
\end{align} 

For $J^{\rm pol}$, the function ${\cal S}(J^{\rm pol})$ is decreasingly monotonic with degree of polarisation.

\section{Equivalent states and equivalent beams}
\label{Equivalent}
This section contains a central result of the paper. In this section, we address the question: whether classically entangled beams can themselves be mimicked by classically separable beams? As mentioned in section (\ref{Introduction}), by classically separable beams, we mean those beams that can be written as incoherent superpositions of pure separable beams. To answer this question, we introduce  a concept of equivalent beams. This, in turn, is based on the concept of equivalent states, introduced in \cite{Bharath14}. Equivalent separable beams of classically entangled beams generally exist in higher dimensions and can be highly mixed.  We shall show that the equivalent separable beams can be written as incoherent superpositions of $SU(2)$ coherent beams. Experimental preparation of equivalent beams of a $2\times 2$ classically entangled beam is hopefully  feasible because $SU(2)$ coherent beams have already been experimentally realised in the OAM domain \cite{tuan2018realizing}. Interestingly, it is an area in which a quantum mechanical concept enriches the understanding in the classical domain. 

We start with a brief recapitulation of equivalent states with the examples of single qubit and two-qubit states. For a detailed discussion, one can refer to \cite{Bharath14}. For a discussion of separable equivalent states of higher dimensional entangled states, we refer the reader to \cite{Adhikary16}.
\subsection{Q-representation and equivalent states in the quantum domain}

\subsubsection{Q--representation of a quantum state}
In quantum mechanics,  a $SU(2)$--coherent state $|\hat{n}(\alpha, \beta)\rangle$ (originally introduced for spin systems) is generated by operation of the $(2T+1)$-- dimensional irreducible representation of the group,  $SU(2)$, on the state $|T_3=+T\rangle$ \cite{Radcliffe71}. That is to say,
\begin{align}
\label{coherent}
|\hat{n}(\alpha, \beta)\rangle = e^{-iT_3\beta}e^{-iT_2\alpha}e^{-iT_3\gamma}|T_3=+T\rangle,
\end{align} 
where  $T_2$  and $T_3$ are two of the generators of $SU(2)$ ($T_1$ being the third one, which is not required in the definition of $SU(2)$ coherent states). $\gamma$ is an uninteresting overall phase. The set of $SU(2)$-coherent states $\{|\hat{n}(\alpha, \beta)\rangle\}$ forms an overcomplete set, i.e.,
\begin{align}
\dfrac{2T+1}{4\pi}\int \sin\alpha d\alpha d\beta |\hat{n}(\alpha, \beta)\rangle\langle \hat{n}(\alpha, \beta)| =\mathbb{1}.
\end{align}
Thus, a density matrix $\rho$ can be expanded in this overcomplete basis. Its Q--function, denoted by $F(\hat{n})$, is defined as,
\begin{align}
F(\hat{n}) =\frac{2T+1}{4\pi} \langle \hat{n}(\alpha, \beta)|\rho|\hat{n}(\alpha, \beta)\rangle.
\end{align}
Due to the overcompleteness property of $SU(2)$--coherent states, $F(\hat{n})$ completely determines the density matrix $\rho$. 
\subsubsection{Equivalent states}
Let there be two states $\rho_1$ and $\rho_2$, belonging to the Hilbert spaces, ${\cal H}^{d_1}$ and ${\cal H}^{d_2}$, of dimensions $d_1$ and $d_2$ respectively. The states $\rho_1 $ and $\rho_2$ are termed as {\it equivalent} if they share the same Q--representation \cite{Bharath14}. For example, a qubit density matrix $\rho^{1/2}(\vec{p}) = \dfrac{1}{2}(\mathbb{1}+\vec{\sigma}\cdot\vec{p})$ has the following $(2T+1)$--dimensional equivalent,
\begin{align}
\label{family}
\rho^T(\vec{p}) = \dfrac{1}{2T+1}(\mathbb{1}+\hat{T}\cdot\vec{p}),~\hat{T}=\dfrac{\vec{T}}{T}.
\end{align}
Evidently, states equivalent to a pure qubit state are all mixed. It is obvious from equation (\ref{family}) that, for $T=\frac{1}{2}$ and $|\vec{p}|=1$, the state $\rho^{1/2}(\vec{p})$ is pure, whereas for all higher $T$ values,  it is mixed.  The crucial point is that the family of states $\rho^T(\vec{p})$ have one and the same $Q$--function, given by,
\begin{align}
F(\hat{n}) = \dfrac{1}{4\pi}(1+\vec{p}\cdot\hat{n}).
\end{align}
Equivalent states have the same information content and hence, they are equally good resources for information processing. In order to extract that information, the concept of equivalent observables is introduced below.  
\subsubsection{Equivalent observables}
Let $\rho_1 \in {\cal H}^{d_1}$ and $\rho_2 \in {\cal H}^{d_2}$ be two equivalent states. The observables $\hat{O}\in {\cal H}^{d_1}$ and $\hat{O}'\in {\cal H}^{d_2}$ are termed as equivalent observables if,
\begin{align}
{\rm Tr}(\rho_1\hat{O}) = {\rm Tr}(\rho_2\hat{O}').
\end{align}

For example, the observable $\hat{O} \equiv \vec{\sigma}\cdot\hat{m}$ for a qubit has its equivalent observable $\hat{O}'\equiv \frac{3T}{T+1}\hat{T}\cdot\hat{m}$ belonging to a $(2T+1)$--dimensional space.

We now turn our attention to  two-qubit systems. The most general density matrix of a two-qubit system is written as,
\begin{align}
\rho^{\frac{1}{2}, \frac{1}{2}} =\dfrac{1}{4}\big[\mathbb{1}+\vec{\sigma}^A\cdot\vec{p}+\vec{\sigma}^B\cdot\vec{q}+t_{ij}\sigma^A_{i}\sigma^B_{j}\big],
\end{align}
 where, the parameters $p_i, q_i$ and $t_{ij}$ characterise the state. While $\vec{p}$ and $\vec{q}$ correspond to local terms of subsystems $A$ and $B$ respectively, $t_{ij}$ represents  correlations between the two subsystems.  The $\frac{1}{2}\otimes T$-- equivalent state  of $\rho^{\frac{1}{2}, \frac{1}{2}}$ is given by,
\begin{align}
\rho^{\frac{1}{2}, T} =\dfrac{1}{2(2T+1)}\big[\mathbb{1}+\vec{\sigma}^A\cdot\vec{p}+\hat{T}^B\cdot\vec{q}+t_{ij}\sigma_{i}^A\hat{T}_{j}^B\big], \hat{T}^B = \dfrac{\vec{T}^B}{T}.
\end{align}
 The family of states $\rho^{\frac{1}{2}, T}$ has the same Q--representation, which is as follows:
\begin{align}
F(\hat{m}, \hat{n}) = \dfrac{1}{(4\pi)^2}\big[1+\hat{m}\cdot\vec{p}+\hat{n}\cdot\vec{q}+t_{ij}m_{i}n_{j}\big].
\end{align} 
For the special case of two-qubit Werner states,
\begin{align}
\rho_W^{\frac{1}{2}, \frac{1}{2}}[\eta] = \dfrac{1}{4}(\mathbb{1}-\eta\vec{\sigma}^A\cdot\vec{\sigma}^B);
  ~\eta \in \Big[-\frac{1}{3}, 1\Big],
\end{align}
the equivalent $\frac{1}{2}\otimes T$ states are given by,
\begin{align}
\label{Equivalent_sep}
\rho_W^{\frac{1}{2}, T}[\eta] = \dfrac{1}{2(2T+1)}(\mathbb{1}-\eta\vec{\sigma}^A\cdot\hat{T}^B),~\eta \in \Big[-\frac{T}{T+1}, 1\Big].
\end{align}
Crucially, the state $\rho_W^{\frac{1}{2}, T}[\eta]$ is separable in the range,
\begin{align}
\label{Smin}
|\eta| \leq \frac{T}{T+1}.
\end{align}
It implies that a lower dimensional entangled state can have higher dimensional separable states as its equivalents. This  feature has been termed as `classical simulation of entangled states'. The minimum value of $T$, for which the $\frac{1}{2}\otimes T$ separable equivalents of two-qubit Werner states exist, can be obtained from equation (\ref{Smin}). If we denote the minimum value of $T$ by $T_{\rm min}$, it is given by the minimum half-integer  greater than $\dfrac{|\eta|}{1-|\eta|}$. We consider a few values of $\eta$ and find out the corresponding $T_{\rm min}$ as follows:
\begin{enumerate}
\item For $\eta = 0.5$, $T_{\rm min}=1$, i.e., the equivalent separable state belongs to a three-dimensional Hilbert space.
\item For $\eta=1$, i.e., for the pure Bell state, the equivalent separable state belongs to the infinite dimensional Hilbert space. 
\end{enumerate}

 So, for $\eta \neq 1$, separable equivalent states belong to  finite dimensional Hilbert spaces.
 The equivalent higher dimensional separable states yield, in principle, the same advantage as the lower dimensional entangled states. For this reason, they can be harnessed in information processing tasks. Some applications of the equivalent separable states in quantum communication protocols have been shown in \cite{Asthana21}. 
 
 The degree of mixedness of the family of states, given in (\ref{Equivalent_sep}), is given by the quantity $1-{\rm Tr}\Big(\rho_W^{\frac{1}{2}, T}[\eta]\Big)^2$, which in terms of $\eta$ and $T$ is given as,
\begin{align}
{\cal M} \equiv 1-{\rm Tr}\Big(\rho_W^{\frac{1}{2}, T}[\eta]\Big)^2= \dfrac{1}{4(2T+1)}\Big(4(2T+1)-2-2\eta^2\dfrac{T+1}{T}\Big).
\end{align}
So, for a given value of $\eta$, the degree of mixedness ${\cal M}$ increases as $T$ increases. It is reflected by the derivative,
\begin{align}
\dfrac{d{\cal M}}{dT}=\dfrac{\eta^2(2T^2+4T+1)}{2T^2(2T+1)^2}+\dfrac{1}{(2T+1)^2},
\end{align}
 which is always positive. So, information processing with equivalent separable states yields a two-fold advantage: it allows for information processing with mixed states, which are separable as well. Of course, the proposed protocols do not yield 100\% fidelity as the equivalent separable state of a pure Bell state belongs to the infinite dimensional Hilbert space.
\subsection{Q--representation and equivalent beams in the classical domain}
We are in a position to advance the concept of equivalent beams. Different degrees of freedoms of optical beams, e.g., polarisation and OAM form Hilbert spaces.  So, the concept of equivalence is amenable to light beams as well. 
\subsubsection{$SU(2)$ coherent optical beams}
\label{SU(2)_coherent_optical_beams}
 The classical counterpart of a $SU(2)$ coherent state is a $SU(2)$ coherent optical beam, defined by \cite{tuan2018realizing},
\begin{align}
\label{Coherent_beam}
\big|\hat{n}(\alpha, \beta)\big) = e^{-iT_3\beta}e^{-iT_2\alpha}e^{-iT_3\gamma}\big|T_3=+T\big).
\end{align} 
  As in equation (\ref{coherent}), $T_2$ and $T_3$ are two of the generators of $SU(2)$ ($T_1$ being the third one, which is not required in the definition of $SU(2)$ coherent beams). $\gamma$ is an uninteresting overall phase. $T_2$ and $T_3$  are  defined with  respect to  a convenient basis consisting of $(2T+1)$ orthonormal modes of electromagnetic beams. The physical significance of $T_1, T_2, T_3$ depends on the context. The beam $|T_3=+T)$ corresponds to the eigenfunction of $T_3$ with the highest eigenvalue and  the unit vector $\hat{n}$ is given by,
\begin{align}
    \hat{n}\equiv \sin\alpha\cos\beta~\hat{e}_1+\sin\alpha\sin\beta~\hat{e}_2+\cos\alpha~\hat{e}_3,
\end{align}
where $\hat{e}_1, \hat{e}_2$ and $\hat{e}_3$ are mutually orthonormal unit vectors.

As an illustration, if we take a three-dimesnional Hilbert space spanned by three Laguerre Gauss modes $|{\rm LG}_{0-1}), |{\rm LG}_{00}), |{\rm LG}_{01})$ and identify them with repective eigenstates of $T_3$, the matrix representations of $T_1$ and $T_2$ have the following forms:
\begin{align}
{T}_1 &=
\begin{blockarray}{cccc}
 & |{\rm LG}_{0-1}) & |{\rm LG}_{00}) & |{\rm LG}_{01})  \\
\begin{block}{c(ccc)}
  ( {\rm LG}_{0-1}| & 0 & \frac{1}{\sqrt{2}} & 0  \\
  ( {\rm LG}_{00}| &  \frac{1}{\sqrt{2}} & 0 &  \frac{1}{\sqrt{2}}  \\
  ( {\rm LG}_{01}| & 0 &  \frac{1}{\sqrt{2}} & 0  \\
\end{block}
\end{blockarray}\nonumber
\end{align}
\begin{align}
\label{OAMSpinmatrices}
{T}_2 &=\begin{blockarray}{cccc}
 & |{\rm LG}_{0-1}) & |{\rm LG}_{00}) & |{\rm LG}_{01})  \\
\begin{block}{c(ccc)}
  ({\rm LG}_{0-1}| & 0 & - \frac{i}{\sqrt{2}} & 0  \\
  ({\rm LG}_{00}| &  \frac{i}{\sqrt{2}} & 0 & - \frac{i}{\sqrt{2}}  \\
  ({\rm LG}_{01}| & 0 &  \frac{i}{\sqrt{2}} & 0  \\
\end{block}
\end{blockarray}.
\end{align}
The $SU(2)$-coherent optical beam $\big|\hat{n}(\alpha, \beta)\big)$ would have the following form,
\begin{align}
\label{Coherent_LG_state}
\big|\hat{n}(\alpha, \beta)\big) &= e^{-iT_3\beta}e^{-iT_2\alpha}\big|{\rm LG}_{01}\big)\nonumber\\
&=\dfrac{1}{2}e^{i\beta}(1-\cos\alpha)|{\rm LG}_{0-1})-\dfrac{\sin\alpha}{\sqrt{2}}|{\rm LG}_{00})\nonumber\\
&+\dfrac{1}{2}e^{-i\beta}(1+\cos\alpha)|{\rm LG}_{01}).
\end{align}
We note that the physical significance of $T_1, T_2$ and $T_3$ depends on the basis chosen and the manner in which they are ordered.

Projecting the beam (\ref{Coherent_LG_state}) onto the $xy$--plane (as is conventional, we have taken $z=0$) in terms of polar coordinates $(r, \phi)$, 
\begin{align}
\big(r, \phi\big|\hat{n}(\alpha, \beta)\big)&=\dfrac{1}{2}e^{i\beta}(1-\cos\alpha){\rm LG}_{0-1}(r,\phi)-\dfrac{\sin\alpha}{\sqrt{2}}{\rm LG}_{00}(r,\phi)\nonumber\\
&+\dfrac{1}{2}e^{-i\beta}(1+\cos\alpha){\rm LG}_{01}(r,\phi),
\end{align}
where the expressions of $LG_{pl}(r,\phi)$ are standard and can be found in, e.g.,  \cite{andrews2012angular}.

 The intensity of the coherent beam $|\hat{n}(\alpha, \beta)\big)$ is given by,
\begin{align}
    I_{\alpha,\beta}(r,\phi)=\big\vert\big(r,\phi\big|\hat{n}(\alpha, \beta)\big)\big\vert^2.
\end{align}

\begin{figure}[htbp]
\centerline{\includegraphics[scale=0.13]{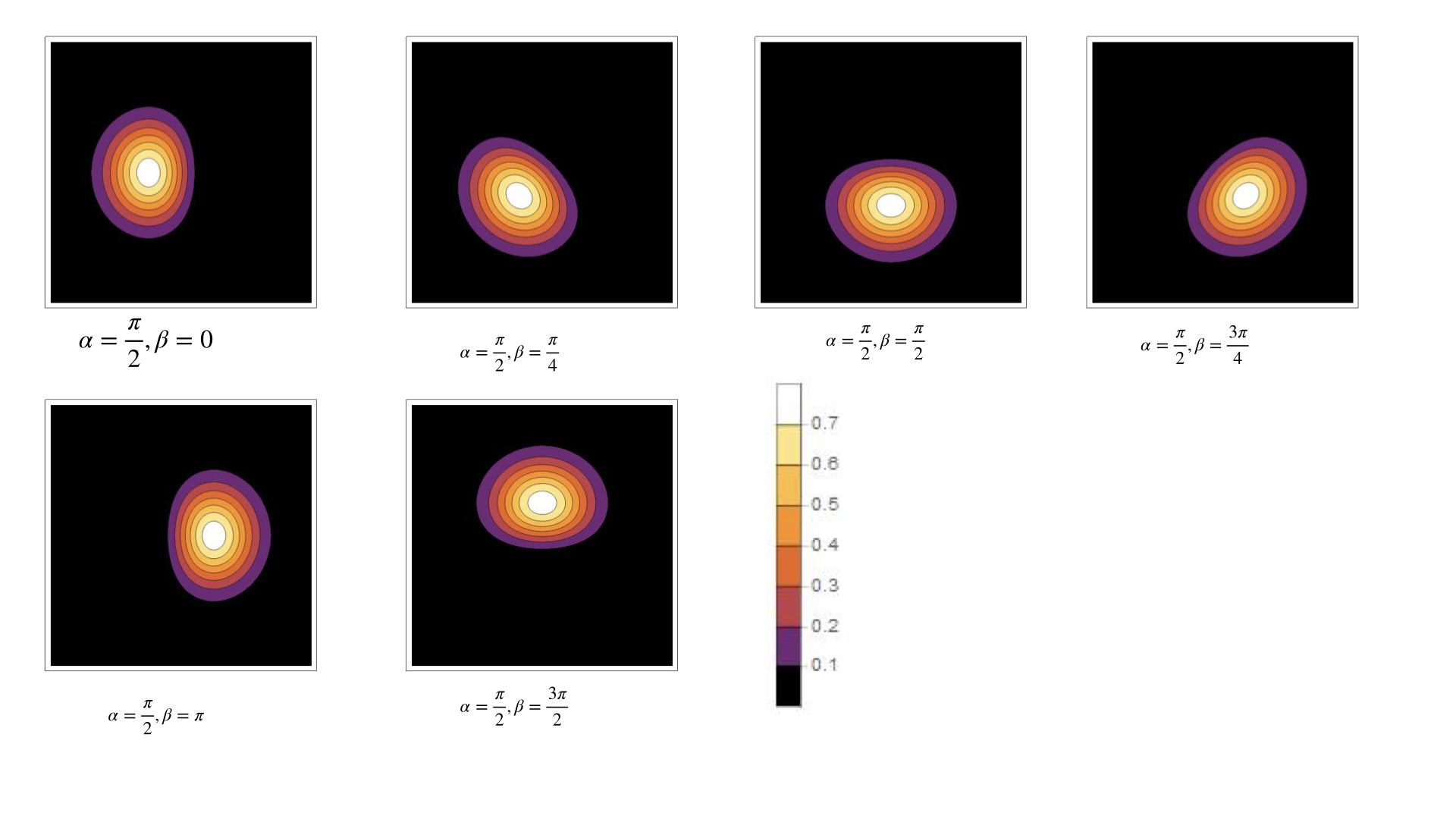}}
\caption{Intensity profile of $SU(2)$ coherent beam given in equation (\ref{Coherent_LG_state}) corresponding to $\alpha=\frac{\pi}{2}$ and different values of $\beta$,  i.e.,  plot of $I_{\frac{\pi}{2}, \beta}(r, \phi)$ for different values of $\beta$}.
\label{Intensity2}
\end{figure}

\begin{figure}[htbp]
\centerline{\includegraphics[scale=0.13]{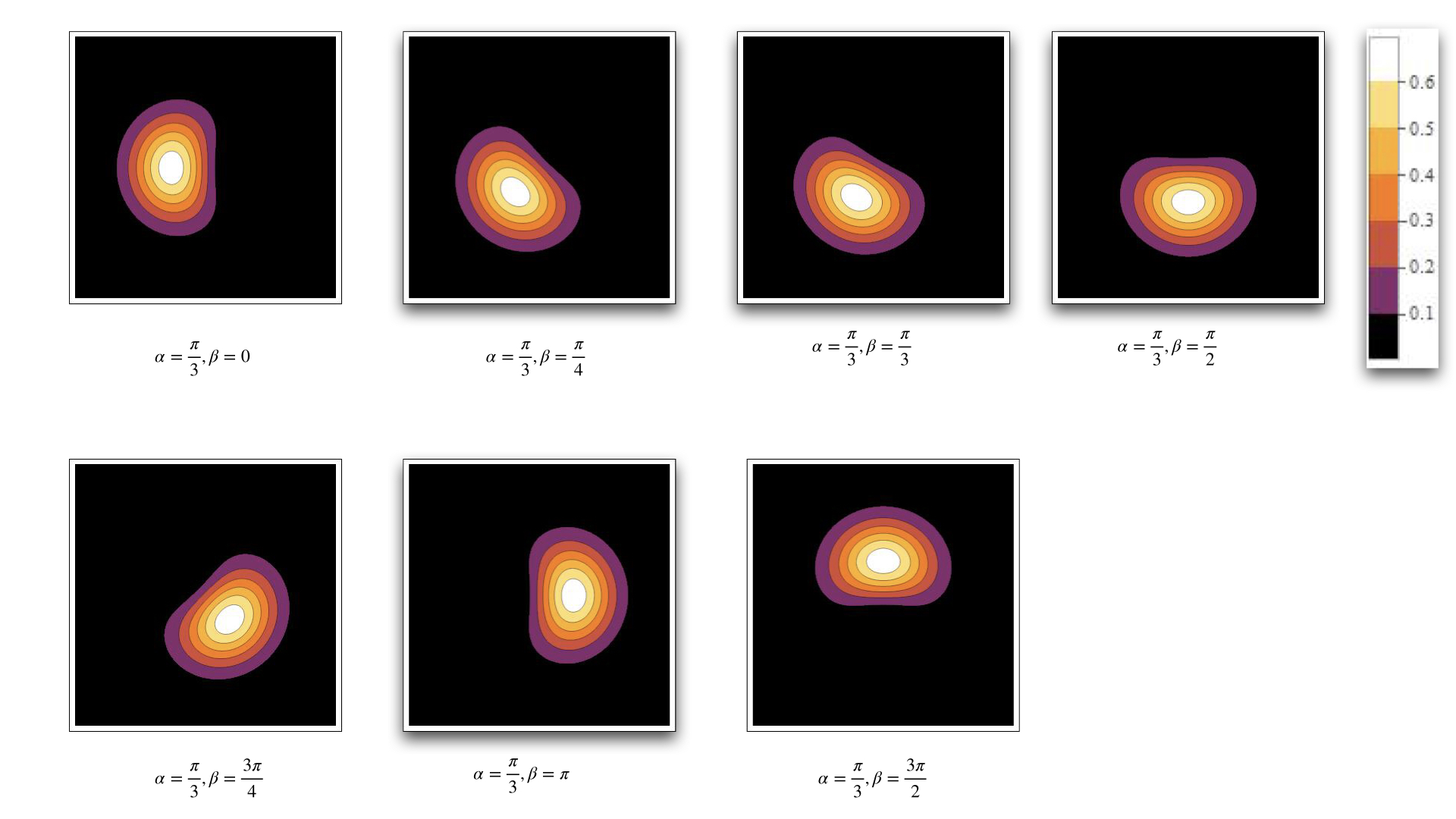}}
\caption{Intensity profile of $SU(2)$ coherent beam  given in equation (\ref{Coherent_LG_state})} corresponding to $\alpha=\frac{\pi}{3}$ and different values of $\beta$,  i.e.,  plot of $I_{\frac{\pi}{3}, \beta}(r, \phi)$ for different values of $\beta$.
\label{Intensity3}
\end{figure}

\begin{figure}[htbp]
\centerline{\includegraphics[scale=0.13]{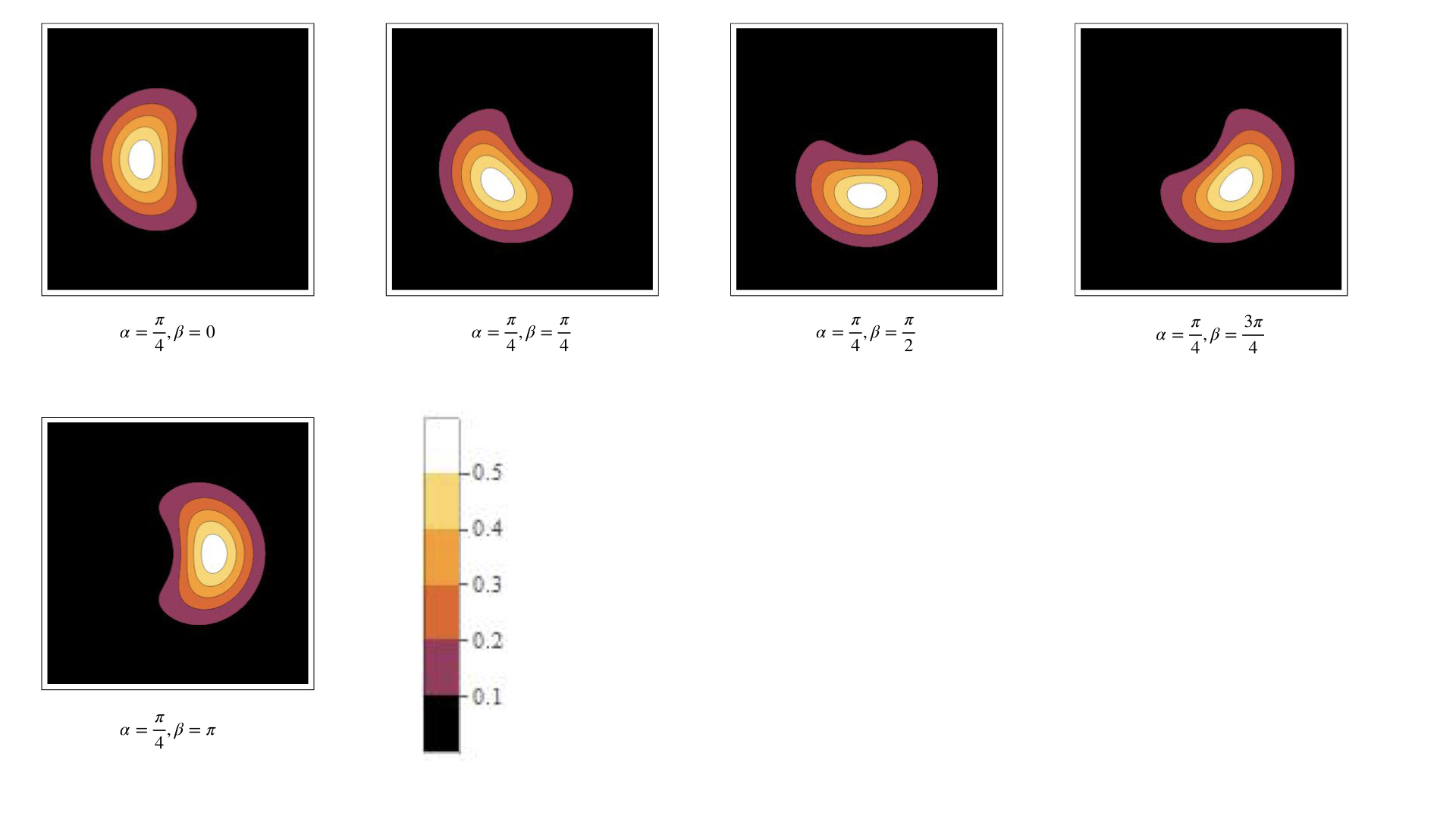}}
\caption{Intensity profile of $SU(2)$ coherent beam  given in equation (\ref{Coherent_LG_state})} corresponding to $\alpha=\frac{\pi}{4}$ and different values of $\beta$,  i.e.,  plot of $I_{\frac{\pi}{4}, \beta}(r, \phi)$ for different values of $\beta$.
\label{Intensity4}
\end{figure}

It is of great interest and importance for us to see the  transverse intensity profiles $I_{\alpha,\beta}(r, \phi)$ of these beams. We show some characteristic profiles of the beams $\big|\hat{n}(\alpha, \beta)\big)$ for $\alpha = \frac{\pi}{4}, \frac{\pi}{3}, \frac{\pi}{2}$ and $\beta=0, \frac{\pi}{4}, \frac{\pi}{2}, \frac{3\pi}{4}, \pi$, in transverse planes, in figures (\ref{Intensity2}), (\ref{Intensity3}) and (\ref{Intensity4}).

Similarly, $SU(2)$--coherent optical beams $\big|\hat{n}(\alpha, \beta)\big)$ can be constructed in higher dimensions. These beams have been realised experimentally upto eight dimensions in \cite{tuan2018realizing}, by employing astigmatic mode transformations of geometric modes  generated by Nd-YAG laser.  

\subsubsection{Q--representation of an optical beam}
The set of $SU(2)$-coherent beams $\{\big|\hat{n}(\alpha, \beta)\big)\}$ forms an overcomplete set, i.e.,
\begin{align}
\dfrac{2T+1}{4\pi}\int \sin\alpha d\alpha d\beta \big|\hat{n}(\alpha, \beta)\big)\big( \hat{n}(\alpha, \beta)\big| =\mathbb{1}.
\end{align}
Thus, any optical beam, whose coherent mode representation is given by a matrix $J$, can be expanded in this overcomplete basis provided by $\{\big|\hat{n} (\alpha, \beta)\big)\}$. The  corresponding Q--function, denoted by $F(\hat{n})$, is simply defined as,
\begin{align}
F(\hat{n}) =\frac{2T+1}{4\pi} \big(\hat{n}(\alpha, \beta)\big|J\big|\hat{n}(\alpha, \beta)\big).
\end{align}
Due to overcompleteness, the diagonal elements $\big( \hat{n}(\alpha, \beta)|J\big|\hat{n}(\alpha, \beta)\big)$ completely determine the matrix $J$. Thus, the complete information of the beam is mapped to a single function $F(\hat{n})$, defined on a sphere.
\subsubsection{Equivalent beams}
Let $J_1$ and $J_2$ be  coherent mode representations of two beams (or of two degrees of freedom of the same beam) belonging to Hilbert spaces of dimensions $d_1$ and $d_2$ respectively. The two beams are termed as {\it equivalent} if they share the same Q--representation. For example, a  polarisation matrix $J^{1/2}(\vec{p}) = \dfrac{1}{2}\Big(\mathbb{1}+\vec{\sigma}\cdot\vec{p}\Big)$ has the following $(2T+1)$-- dimensional equivalent beams whose coherent mode representations are given by,
\begin{align}
\label{family_beam}
J^T(\vec{p}) = \dfrac{1}{2T+1}(\mathbb{1}+\hat{T}\cdot\vec{p}),~\hat{T}=\dfrac{\vec{T}}{T}.
\end{align}
Interestingly, completely polarised beams have higher dimensional mixed beams as their equivalents. The Q-representation of the beams $J^T(\vec{p})$ is given by $F(\hat{n}) = \dfrac{1}{4\pi}\Big(1+\vec{p}\cdot\hat{n}\Big)$.

\begin{figure}[htbp]
\centerline{\includegraphics[scale=0.15]{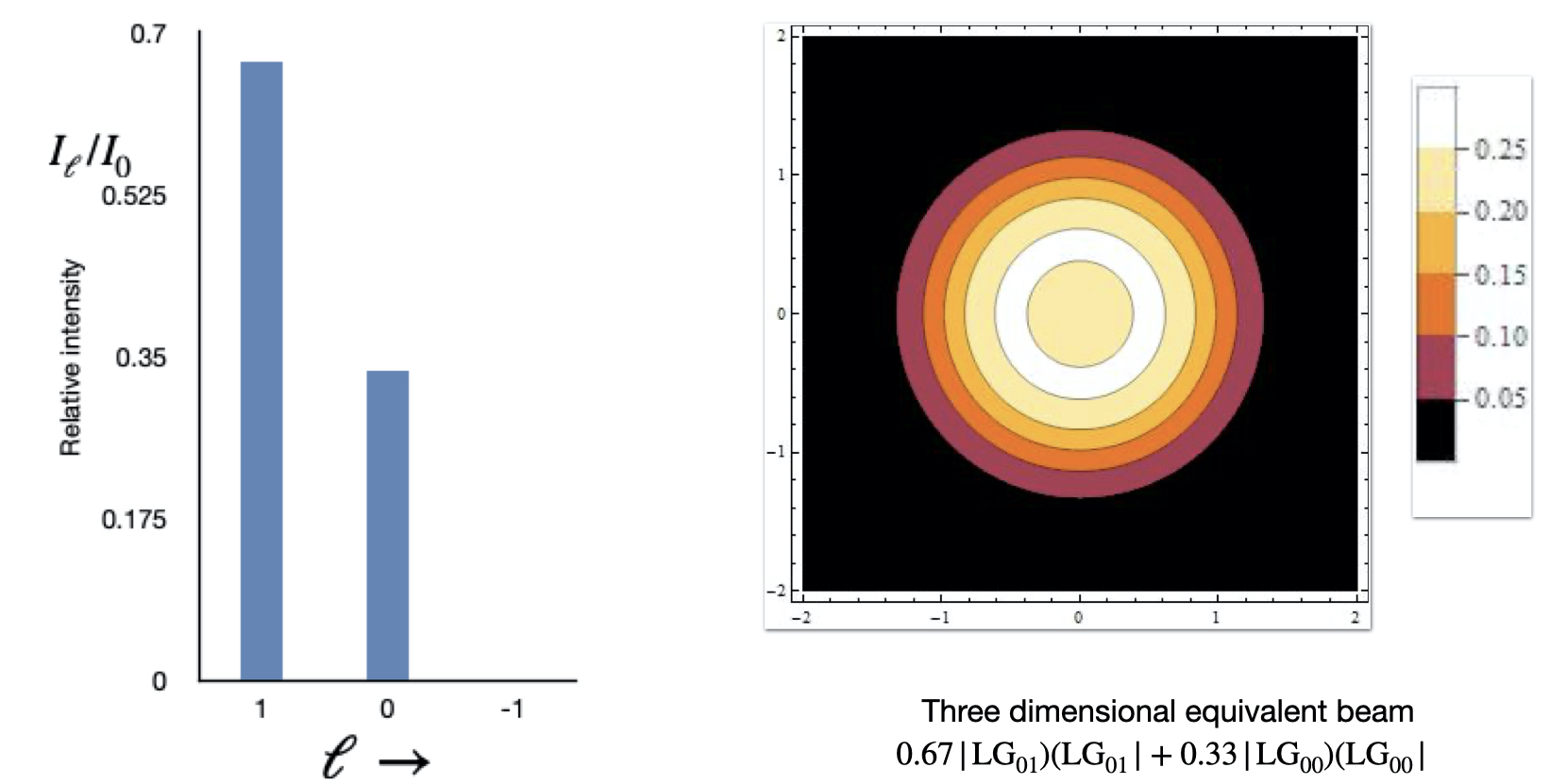}}
\caption{Spectral decomposition of a three dimensional equivalent beam of a pure two dimensional beam. For example, the three dimensional beam may be an OAM beam belonging to the Hilbert space spanned by $\{|{\rm LG}_{0-1}), |{\rm LG}_{00}), |{\rm LG}_{01})\}$ and the two dimensional pure beam may correspond to a fully polarised beam.  The transverse intensity profile of the beam is shown on the right. $\ell=0$ and $\ell=1$  contribute approximately $35\%$ and $65\%$ respectively to intensity. Note that $I_0$ corresponds to the total intensity.}
\label{l=3}
\end{figure}

\begin{figure}[htbp]
\centerline{\includegraphics[scale=0.15]{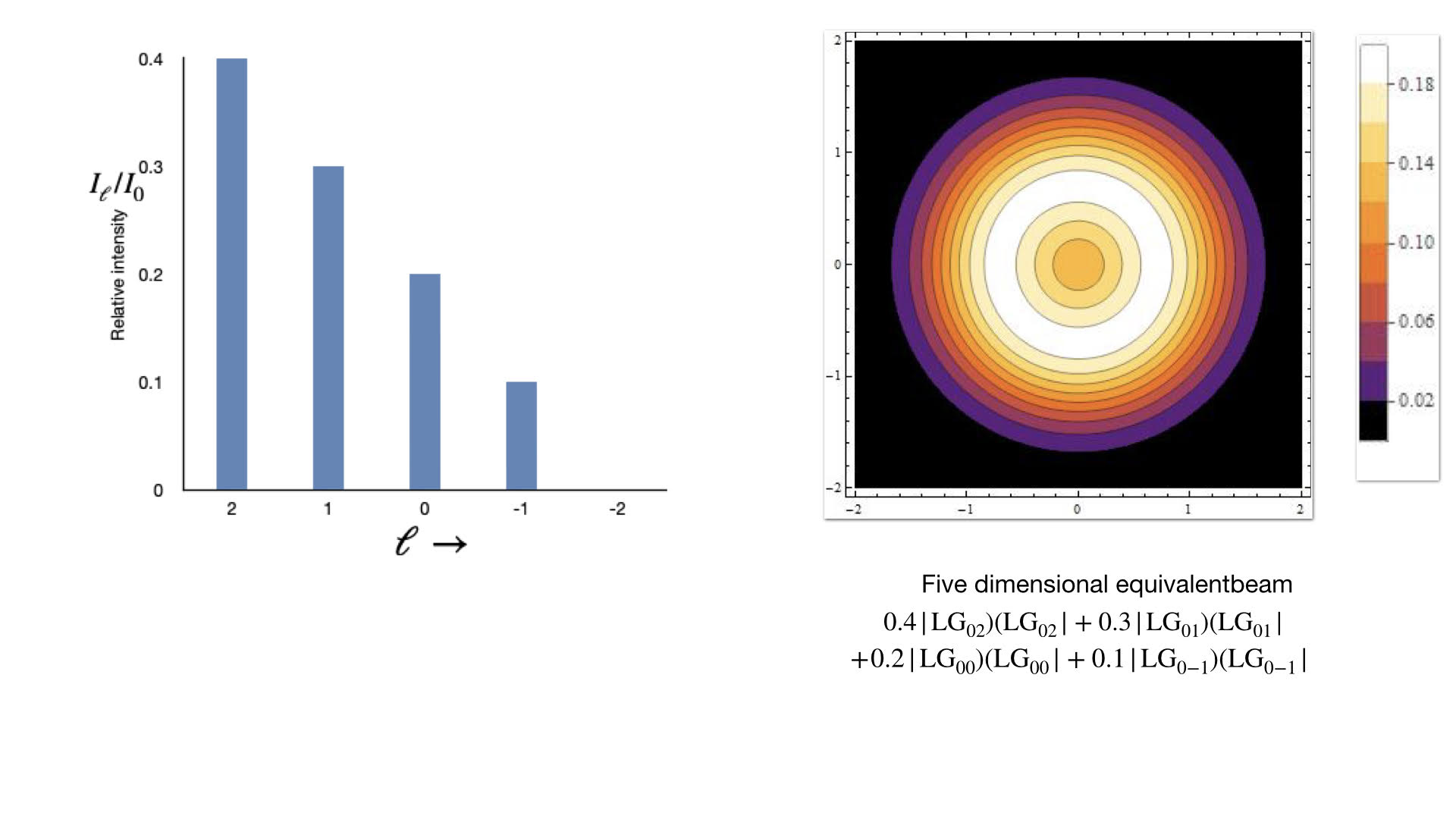}}
  \caption{Spectral decomposition of a five dimensional equivalent beam of a pure two dimensional beam. For example, the five dimensional beam may be an OAM beam belonging to the Hilbert space spanned by $\{|{\rm LG}_{0-2}), |{\rm LG}_{0-1}), |{\rm LG}_{00}), |{\rm LG}_{01}), |{\rm LG}_{02})\}$ and the two dimensional pure beam may correspond to a fully polarised beam.   The transverse intensity profile of the beam is shown on the right. $\ell=2, 1, 0$ and $-1$  contribute $40\%$, $30\%, 20\%$ and $10\%$ respectively to intensity. Note that $I_0$ corresponds to the total intensity.} 
  \label{l=5}
\end{figure}

The spectral decompositions of equivalent OAM beams of fully polarised light with the choice, $\vec{p}=\begin{pmatrix}
0, ~0, ~1
\end{pmatrix}^T$,  belonging to three and five dimensional Hilbert spaces, are shown in figures (\ref{l=3}) and (\ref{l=5}) respectively.  The normalised intensity, for the case of three--dimensional equivalent beam ($T=1$) in the transverse plane ($z=0$) is given by,
\begin{align}
I^{(1)}(r, \phi)=0.67|{\rm LG}_{01}(r,\phi)|^2+0.33|{\rm LG}_{00}(r, \phi)|^2.
\end{align}
Similarly, the normalised intensity, for the choice of  five--dimensional equivalent beam ($T=2$) in the transverse plane ($z=0$) is given by,
\begin{align}
    I^{(2)}(r, \phi) &= 0.4|{\rm LG}_{02}(r,\phi)|^2+0.3|{\rm LG}_{01}(r,\phi)|^2\nonumber\\
    &+0.2|{\rm LG}_{00}(r,\phi)|^2+0.1|{\rm LG}_{0-1}(r,\phi)|^2.
\end{align}
We have also shown the intensity profiles of the corresponding OAM beams  in transverse planes, that can be observed in experiments, in figures (\ref{l=3}) and (\ref{l=5}).
\subsubsection{Equivalent observables for equivalent beams}
Equivalent observables for equivalent beams can be defined in a similar way, as in the quantum case, with the sole provision that probability maps to the normalised intensity. As an example, the two-dimesnional  observable $\vec{\sigma}\cdot\vec{m}$ maps to the observable $\frac{3T}{T+1}\hat{T}\cdot\vec{m}$ in $(2T+1)$ dimensions. 

As an illustration, let a three-dimensional Hilbert space be spanned by the modes $\{|{\rm LG}_{0-1}), |{\rm LG}_{00}), |{\rm LG}_{01}))\}$. For the particular choice of,
\begin{align}
\hat{m} = \sin\lambda\cos\mu~\hat{e}_1+\sin\lambda\sin\mu~\hat{e}_2+\cos\lambda~\hat{e}_3,~~|\hat{e}_1\cdot(\hat{e}_2\times \hat{e}_3)|=1,
\end{align}
 the matrix representation of the  observable $\frac{3T}{T+1}\hat{T}\cdot\vec{m}$ in this Hilbert space is given by,
\begin{align}
\frac{3T}{T+1}\hat{T}\cdot\hat{m}\Big\vert_{T=1} &=
\dfrac{3}{2}\begin{pmatrix}
    -\cos\lambda & \frac{\sin\lambda e^{-i\mu}}{\sqrt{2}} & 0  \\
  \frac{\sin\lambda e^{i\mu}}{\sqrt{2}} & 0 & \frac{\sin\lambda e^{-i\mu}}{\sqrt{2}}  \\
  0 & \frac{\sin\lambda e^{i\mu}}{\sqrt{2}} & \cos\lambda  \\
\end{pmatrix}.
\label{Observable_explicit}
\end{align}
 The eigen-resolution of the observable, given in (\ref{Observable_explicit}), is as follows:
\begin{align}
\frac{3T}{T+1}\hat{T}\cdot\vec{m}\vert_{T=1}=\frac{3}{2}\hat{T}\cdot\hat{m}=\dfrac{3}{2}\big(|v_1)(v_1|-|v_2)(v_2|\big),
\end{align}
where,
\begin{align}
|v_1)(v_1|=&\dfrac{1}{2}\vec{T}\cdot\hat{m}(\vec{T}\cdot\hat{m}+\mathbb{1})\nonumber\\
 |v_2)(v_2|=&\dfrac{1}{2}\vec{T}\cdot\hat{m}(\vec{T}\cdot\hat{m}-\mathbb{1})\ ~{\rm and},\nonumber\\
 |v_3)(v_3|=&\mathbb{1}-(\vec{T}\cdot\hat{m})^2.
\end{align}
So, a measurement of the observable $\frac{3}{2}\vec{T}\cdot\hat{m}$ is tantamount to measuring intensities corresponding to three one-dimensional projections $|v_1)(v_1|$, $|v_2)(v_2|$ and $|v_3)(v_3|$.

\subsubsection{Equivalents of classically entangled beams}
\label{Equivalent_of_classically_entangled_beams}
Experimental generation of a classically entangled beam has become quite common (see, for example \cite{Forbes19} and references therein), e.g., the beams entangled in polarisation and OAM degrees of freedom. So, in this section, we study the properties of equivalent beams of an optical beam mimicking a two-qubit system. 

Let the orthonormal bases of two degrees of freedom of a beam be $\{|0)_A, |1)_A\}$ and $\{|0)_B, |1)_B\}$ respectively.
 The coherent mode representation of the most general beam  in these two degrees of freedom is written as,
\begin{align}
J^{\frac{1}{2}, \frac{1}{2}} =\dfrac{1}{4}\big[\mathbb{1}+\vec{\sigma}^A\cdot\vec{p}+\vec{\sigma}^B\cdot\vec{q}+t_{ij}\sigma_{i}^A\sigma_{j}^B\big], i, j\in\{1,2,3\},
\end{align}
where $\sigma_{i}^A$ and $\sigma_{j}^B$ represent the analogues of Pauli observables in the two degrees of freedom. The coherent mode representation of its $\frac{1}{2}\otimes T$-- equivalent beam is given by,
\begin{align}
J^{\frac{1}{2}, T} =\dfrac{1}{2(2T+1)}\big[\mathbb{1}+\vec{\sigma}^A\cdot\vec{p}+\hat{T}^B\cdot\vec{q}+t_{ij}\sigma_{i}^A\hat{T}_{j}^B\big], \hat{T}^B = \dfrac{\vec{T}^B}{T}.
\end{align}
The family of beams $J^{\frac{1}{2}, T}$ has the same Q--representation, which is as follows:
\begin{align}
F(\hat{m}, \hat{n}) = \dfrac{1}{(4\pi)^2}\big[1+\hat{m}\cdot\vec{p}+\hat{n}\cdot\vec{q}+t_{ij}m_{i}n_{j}\big].
\end{align} 
Of special interest to us is the beam, whose coherent mode representation has the same structure as Werner--states in quantum mechanics, i.e., 
\begin{align}
\label{Wener_beam}
J_W^{\frac{1}{2}, \frac{1}{2}}[\eta] = \dfrac{1}{4}(\mathbb{1}-\eta\vec{\sigma}^A\cdot\vec{\sigma}^B),~\eta\in \Big[-\frac{1}{3}, 1\Big].
\end{align}
We term these beams as {\it Werner beams}. The coherent mode representation of the equivalent $\frac{1}{2}\otimes T$ beams is given by,
\begin{align}
\label{Equivalent_Werner_beam}
J_W^{\frac{1}{2}, T}[\eta] = \dfrac{1}{2(2T+1)}(\mathbb{1}-\eta\vec{\sigma}^A\cdot\hat{T}^B),~ \eta \in \Big[-\frac{T}{T+1}, 1\Big].
\end{align}
 The beam (\ref{Equivalent_Werner_beam}), $J_W^{\frac{1}{2}, T}[\eta]$, is separable in the range $|\eta| \leq \frac{T}{T+1}$.  In equation (\ref{Wener_beam}), all the beams corresponding to $\eta > \frac{1}{3}$ are entangled, however, the beam (\ref{Equivalent_Werner_beam}) is entangled in the range $\frac{T}{T+1}< \eta \leq 1$. Thus, the range of $\eta$, in which the beam (\ref{Equivalent_Werner_beam}) is entangled, shrinks as $T$ increases. It implies that a lower dimensional entangled beam has a higher dimensional separable beam as its equivalent.  We shall use the acronym `SEW' for separable equivalent of $2\times 2$--entangled Werner beam.   Separable equivalent beams can, in principle, be prepared by incoherent superposition of pure separable beams. 

 \subsubsection{Separable expansion of $J_W^{\frac{1}{2}, 1}[\eta]$ for $\eta=0.5$}
 \label{Separable_decomposition_1}
We now consider an example. At $\eta=\frac{1}{2}$, $J^{\frac{1}{2},\frac{1}{2}}[\eta]$ represents a classically entangled beam. Its equivalent beam, for $T=1$, is, however separable. In fact, its separable expansion is given by,
\begin{align}
\label{Separable_expansion}
&\dfrac{1}{6}\Big[\mathbb{1}-\dfrac{1}{2}\vec{\sigma}^A\cdot\vec{T}^B\Big]\nonumber\\
=&\dfrac{1}{6}\Big\{|V)(V|\otimes|\psi_1)(\psi_1| +|H)(H|\otimes|\psi_2)(\psi_2|\nonumber\\
+&|-)(-|\otimes|\psi_{3+})(\psi_{3+}|+|+)(+|\otimes|\psi_{3-})(\psi_{3-}|\nonumber\\
+&|R)(R|\otimes|\psi_{4+})(\psi_{4+}|+|L)(L|\otimes|\psi_{4-})(\psi_{4-}|\Big\},
\end{align}
where, the symbols have the following meaning:
\begin{align}
    |\pm) &= \frac{1}{\sqrt{2}}\{|H)\pm|V)\},\nonumber\\ |R)&=\frac{1}{\sqrt{2}}\{|H)+i|V)\}, |L)=\frac{1}{\sqrt{2}}\{|H)-i|V)\},\nonumber\\
    |\psi_1)&=|{\rm LG}_{01});~|\psi_2)=|LG_{0-1})\nonumber\\
    |\psi_{3\pm})&=\frac{1}{2}|{\rm LG}_{0-1})\pm\frac{1}{\sqrt{2}}|{\rm LG}_{00})+\frac{1}{{2}}|{\rm LG}_{01})\nonumber\\
    |\psi_{4\pm}) &=\frac{1}{{2}}|{\rm LG}_{0-1})\pm \frac{i}{\sqrt{2}}|{\rm LG}_{00})-\frac{1}{{2}}|{\rm LG}_{01}).
\end{align}
Thus, preparation of the SEW,  $J^{\frac{1}{2},1}[\frac{1}{2}]$, in the polarisation-OAM degrees of freedom requires an incoherent mixing of six beams as shown in equation (\ref{Separable_expansion}).  Similar separable expansions can be constructed for   $J^{\frac{1}{2},T}[\eta]$ for higher values of $T$, which are given below.  This brings out the advantage that there is no need for spin--orbit coupling in preparation of the equivalent separable beams, which is otherwise needed for generation of $J^{1/2, 1/2}[\eta]$ for the same value of $\eta$.

\subsubsection{Generation of SEW  of $J_W^{\frac{1}{2}, T}[\eta]$ for large $T$}
\label{Separable_for_large_T}
It might appear that one would require a very large number of beams for generation of SEW for a large value of $T$. However, we show below that a set of six and twelve beams would generate the SEW to an excellent approximation upto $2\times 4$ and $ 2\times 9$ dimensions respectively.  The restriction of a  small number of beams is motivated by experimental considerations, as we cannot afford to have an incoherent superposition of very large number of pure separable beams in polarisation and OAM modes.

Recall that for a given $T$, the beam $J^{\frac{1}{2}, T}_W[\eta]$ is separable upto $\eta_{\rm max} = \frac{T}{T+1}$. Of particular interest to us is the equivalent separable beam $J^{\frac{1}{2}, T}_W[\eta_{\rm max}]$.  This is the beam that simulates the most entangled $2\times 2$ --Werner beam $J_W^{\frac{1}{2}, \frac{1}{2}}[\eta_{\rm max}]$ in $(2T+1)$--dimensions. 

We consider the incoherent superposition of following six beams for preparation of $J^{\frac{1}{2}, T}_W[\eta]$: 
\begin{enumerate}
    \item  the first two beams are eigenmodes of $\sigma^A_1$ and $T^B_1$ with eigenvalues $(+1, -T)$ and $(-1, +T)$,
    \item  the second pair of beams are eigenmodes of $\sigma^A_2$ and $T^B_2$ with eigenvalues $(+1, -T)$ and $(-1, +T)$ respectively, 
    \item the last two beams are eigenmodes of $\sigma_3^A$ and  $T^B_3$ with eigenvalues $(+1, -T)$ and $(-1, +T)$ respectively.
\end{enumerate}

 This choice is motivated by $SU(2)\times SU(2)$--invariance of the SEW. In other words, we consider the following incoherent superposition,
\begin{align}
 J^{(1)}_c\equiv   \frac{1}{6}\sum_{i=1}^3\Big\{& |\sigma^A_i=+1, T^B_i=-T)(\sigma^A_i=+1, T^B_i=-T|\nonumber\\
 \label{Six_beams}
    +&|\sigma^A_i=-1, T^B_i=+T)(\sigma^A_i=-1, T^B_i=+T|\Big\},
\end{align}
where the symbol $|\sigma^A_i=+1, T^B_i=-T)$ represents the simultaneous eigenmode of $\sigma^A_i$ and $T^B_i$ with eigenvalues $+1$ and $-T$ respectively.
 Note that we identify the eigenmodes of the operator $T^B_3$ with the Laguerre--Gauss beams $\{|LG_{0T}), \cdots, |{\rm LG}_{0-T})\}$ and those of $\sigma_3^A$ with $|H)$ and $|V)$. We  find out what the proximity of incoherent superposition of the six beams with the desired beam is. 
    The proximity between the beams $J^{(1)}_c$ and $J_W^{\frac{1}{2}, T}\Big[\frac{T}{T+1}\Big]$ is captured by the norm of,
    \begin{align}
    \label{J1_diff}
        J^{(1)}_{\rm diff}\equiv J^{(1)}_c-J_W^{\frac{1}{2}, T}\Big[\frac{T}{T+1}\Big].
    \end{align}
     We represent the norm by the standard notation $\vert\vert J^{(1)}_{\rm diff}\vert\vert$. The norm are given for different dimensions $2 \times (2T+1)$ in table (\ref{Proximity}). All the pure separable beams given in (\ref{Six_beams}) can be experimentally generated in the polarisation and OAM degrees of freedom by employing spatial light modulators and phase plates.


\begin{center}
\begin{table}[h!]
\begin{tabular}{|| c|c|c|c ||} 
 \hline
Dimension & $||J^{(1)}_{\rm diff}||$ & $||J^{(2)}_{\rm diff}||$ & $||J^{(3)}_{\rm diff}||$  \\ [0.5ex] 
 \hline\hline
 $2\times 3$ & 0 & -- &--\\
 \hline
$2\times 4$ & 0.09 &0.02  & 0.01\\
 \hline
 $2\times 5$ & 0.14 & 0.04  & 0.02\\
 \hline
$2\times 6$ & 0.18 & 0.05 & 0.03\\
 \hline
$2\times 7$ & 0.21 & 0.06  &0.03\\
 \hline
$2\times 8$ & 0.24 & 0.08 & 0.04\\
 \hline
$2\times 9$ & 0.26 &  0.09&  0.05\\
 \hline
$2\times 11$ & 0.29 &0.12 & 0.06 \\
 \hline
$2\times 13$ & 0.3 & 0.14 &  0.08\\
 \hline
$2\times 15$ & 0.32 & 0.16 &0.09\\[1ex] 
 \hline
\end{tabular}
\caption{Proximities of the beam $J^{(1)}_c, J^{(2)}_c$ and $J^{(3)}_c$ with the exact equivalent beam $J_W^{\frac{1}{2}, T}\Big[\frac{T}{T+1}\Big]$, captured by the norms $||J^{(1)}_c||$, $||J^{(2)}_c||$, and $||J^{(3)}||$. The quantities $J^{(1)}_{\rm diff}, J_{\rm diff}^{(2)}$ and $J_{\rm diff}^{(3)}$  have been defined in equations (\ref{J1_diff}) and (\ref{J2_diff})  respectively. }
\label{Proximity}
\end{table}
\end{center}
Since the norm $||J^{(1)}_{\rm diff}||$ increases with increase in dimensions,  we have given two other coherent mode representations $J^{(2)}_c$ and $J^{(3)}_c$ which involve incoherent superpositions of tweleve and eighteen beams respectively in the appendix (\ref{Separable_appendix}). The norm of the differences,
\begin{align}
    & J^{(2)}_{\rm diff}\equiv J^{(2)}_c - J^{\frac{1}{2}, T}_W\Big[\frac{T}{T+1}\Big];\nonumber\\
    \label{J2_diff}
   & J^{(3)}_{\rm diff}\equiv J^{(3)}_c - J^{\frac{1}{2}, T}_W\Big[\frac{T}{T+1}\Big],
\end{align}
  are also given in the table (\ref{Proximity}). It is clear from table (\ref{Proximity}) that increasing the number of beams  in the incoherent superposition to twelve and eighteen may reduce the norm $J^{(2)}_{\rm diff}$ and $J^{(3)}_{\rm diff}$,  roughly by a factor of 5-10.

We now turn our attention to the applications of equivalent separable beams for transferring the information between two degrees of freedom of a beam. 
\section{Application: protocol for transfer of information from one degree of freedom to another}
\label{Remote state preparation unknown qubit}
  In this section, we present a protocol for transfer of information from one degree of freedom of a classical light to another degree of freedom, without using classical entanglement. The information need not be a-priori known.

   In the protocol, the separable beam, having the coherent mode representation,
 \begin{align}
 \label{Channel_1111}
 J^{BC}(\eta)^{1/2, T} \equiv \dfrac{1}{2(2T+1)}(\mathbb{1}-\eta\vec{\sigma}^B\cdot\hat{T}^C),
 \end{align}
 in, say, polarisation and OAM degrees of freedom, acts as a channel. The symbols $B$ and $C$ stand for polarisation and OAM degrees of freedom respectively. Note that the OAM subspace is spanned by $(2T+1)$ OAM modes.  The beam in the equation (\ref{Channel_1111}) can be written as,
\begin{align}
&    \frac{1}{2(2T+1)}(\mathbb{1}-\eta\vec{\sigma}^B\cdot\hat{T}^C) =\nonumber\\ =&(1-\eta)\frac{1}{2(2T+1)}\mathbb{1}+\eta\frac{1}{2(2T+1)}(\mathbb{1}-\vec{\sigma}^B\cdot\hat{T}^C),\nonumber
\end{align}
where the first fraction contributes to the uniform background (noise) and the second fraction (Bell--like beam) contributes to signal.

 The minimum value of $T$, for which the beam given in equation (\ref{Channel_1111}), would be separable, is given by,
 \begin{align}
 \label{Tmineta}
  T_{\rm min}(\eta) \equiv \Bigg[\Big[\dfrac{|\eta|}{1-|\eta|}\Big]\Bigg].
 \end{align}
We employ the symbol $[[x]]$  to represent the closest half--integer greater than or equal to $x$.  Accordingly, the minimum dimension is given by,
\begin{align}
\label{dmin}
   d_{\rm min} = 2T_{\rm min}+1. 
\end{align}
 Clearly,  the minimum dimension $d_{\rm min}$ also increases with increase in $\eta$.

Let $J^A=\frac{1}{2}({1}+\vec{\sigma}^A\cdot\vec{p})$ be the coherent mode representation of an unknown beam in some degree of freedom, say path, whose equivalent is to be prepared (say, in the OAM domain).  The combined coherent mode representation in the three degrees of freedom is given by,
   \begin{equation}
   \label{Complete_beam_2by2T+1}
      {J}^A\otimes J^{BC}(\eta)^{1/2, T}=\frac{1}{2^2(2T+1)}(\mathbb{1}+\vec{\sigma}^A\cdot\vec{p})\otimes(\mathbb{1}-\eta\vec{\sigma}^B\cdot\hat{T}^C).
  \end{equation}
   There is a correlation between polarisation and OAM degrees of freedom, which is reflected in the structure of $J^{BC}(\eta)^{1/2, T}$.  The protocol is as follows:
 \begin{enumerate}
\item  Unlike in quantum mechanics, which is intrinsically probabilistic, the outcome of a measurement in the classical optical domain is deterministic. Hence, the experimenter may simply project\footnote{The Pauli operators in the polarisation domain are defined as follows: $\sigma^B_{1}=|H)(V|+|V)(H|$, $\sigma^B_{2}=-i|H)(V|+i|V)(H|$, and, $\sigma^B_{3}=|H)(H|-|V)(V|$. Similarly, they can be defined in the path degrees of freedom (A), by identifying $|0)_A$ and $|1)_A$ to be orthonormal coherent modes of $\sigma^A_{3}$ with eigenvalues $\pm 1$.} the beam given in (\ref{Complete_beam_2by2T+1}) along $|\psi_1) = \frac{1}{\sqrt{2}}\{|0)_A|V)-|1)_A|H)\}$. Due to this projection, the information of path degree of freedom ($A$) gets transferred to the OAM degree of freedom ($C$) and the intensity of the beam gets reduced to one-fourth of its initial value.  This constitutes a significant simplification over quantum teleportation since no post--processing is required. The beam, after the measurement, in the OAM degree of freedom is described by,
\begin{align}
\label{OAM_final}
    \tilde{J}^C(\eta)\equiv \dfrac{1}{2T+1}(\mathbb{1}+\eta\hat{T}^C\cdot\vec{p}).
\end{align}
 Since the separable equivalent beam of the pure maximally entangled beam (corresponding to $\eta=1$) belongs to the infinite dimensional space, the beam prepared in the OAM degree of freedom is not the exact equivalent of the two-dimensional beam in the path degree of freedom. It can, however, be as close to the exact equivalent beam as desirable by choosing larger and larger values of $T$. 
 \item Should the experimentalist project along any of the other three Bell beams, the intensity will get diminished to one-fourth of its initial value as before. Additionally, it also requires a post-processing. If the experimenter projects along one of the following three Bell beams,
  \begin{eqnarray}
&  |\psi_2)=\frac{1}{\sqrt{2}}\big\{|0)_A|H)-|1)_A|V)\big\};  \nonumber\\ &|\psi_3)=\frac{1}{\sqrt{2}}\big\{|0)_A|H)+|1)_A|V)\big\};\nonumber\\
&|\psi_4)=\frac{1}{\sqrt{2}}\big\{|0)_A|V)+|1)_A|H)\big\},
  \end{eqnarray}
 on the path and polarisation degree of freedom, the transformations to be applied on the OAM degree of freedom are $ R^C_1(\pi), R^C_2(\pi), R^C_3(\pi)$, in the same order. The symbols $R^C_1(\pi), R^C_2(\pi), R^C_3(\pi)$ represent analogue of Wigner rotation matrices in $(2T+1)$-- dimensional Hilbert space about $\hat{e}_1, \hat{e}_2, $ and $\hat{e}_3$ axes through an angle $\pi$ in the OAM domain, i.e., $R^C_{\hat{n}}(\pi)=e^{i\vec{T}^C\cdot\hat{n}\pi}$.

 \end{enumerate}
  The projections along Bell beams can be performed experimentally in the path and polarisation degrees of freedom by employing Sagnac interferometers and a half wave plate. They have already been experimentally  performed in \cite{Guzman16}, which is recapitulated below. If necessary, the unitary transformations (Wigner rotations) on the OAM modes can be performed by using holographic techniques proposed in \cite{Wang17}.  The algebra of the steps underlying the protocol has been worked out in Appendix (\ref{Algebra}).

 \subsection{Experimental setup}
In what follows, we outline an experimental setup that may be used for implementation of this protocol in the polarisation, path and OAM degrees of freedom. It comprises of three parts: (I) generation of a SEW, (II) application of the CNOT gate and projection on the path and polarisation degrees of freedom, and, (III) implementation of unitary transformation in the residual OAM domain. Fortunately, techniques for experimental implementation of (II) and (III) are already available \cite{Guzman16, Wang17}. Hence, we mainly focus on experimental implementation of (I). For the sake of completeness, we also outline how (II) and (III) may be performed. \\
\subsubsection{ Generation of a SEW:} By definition, mixed separable beams can be written as incoherent sums of pure beams. The separable expansion is, however, not unique.  As has been mentioned in the subsections (\ref{Separable_decomposition_1}) and  (\ref{Separable_for_large_T}) of the section (\ref{Equivalent}),  expansions involving only six and twelve beams are sufficient for construction of SEW to an excellent approximation upto $2 \times 4$ and $2 \times 9$ dimensions respectively. We have, however, identified a set of eighteen beams (given in appendix (\ref{Separable_appendix})), whose incoherent superposition is even closer to the actual SEW\footnote{The proximity has been quantified in table (\ref{Proximity}).}. Note that OAM beams with uniform polarisations have been experimentally generated by using spatial light modulators and phase plates (e.g., quarter wave plate and half--wave plate) \cite{Milione12}. The desired beam, having the coherent--mode expansion given in equation (\ref{Channel_1111}), may be hopefully generated by incoherently adding all the beams. \\
\subsubsection{ Generation of a bit in the path degree of freedom:} Following the experimental method adopted in \cite{Guzman16}, we may employ a 50/50 beam
splitter to create two identical copies of
$J^{BC}(\eta)^{T, \frac{1}{2}}$, which then propagate along different paths. The symbol $|0)_A$ and $|1)_A$ represent path 1 and path 2 respectively \cite{Aiello15}. To encode the initial information in each path, two
density filters may be used at each output of the
beam splitter.\\
\subsubsection{Bell--beam projection (via application of CNOT gate):}
\label{Sagnac}
\begin{figure}[htbp]
\centerline{\includegraphics[scale=0.15]{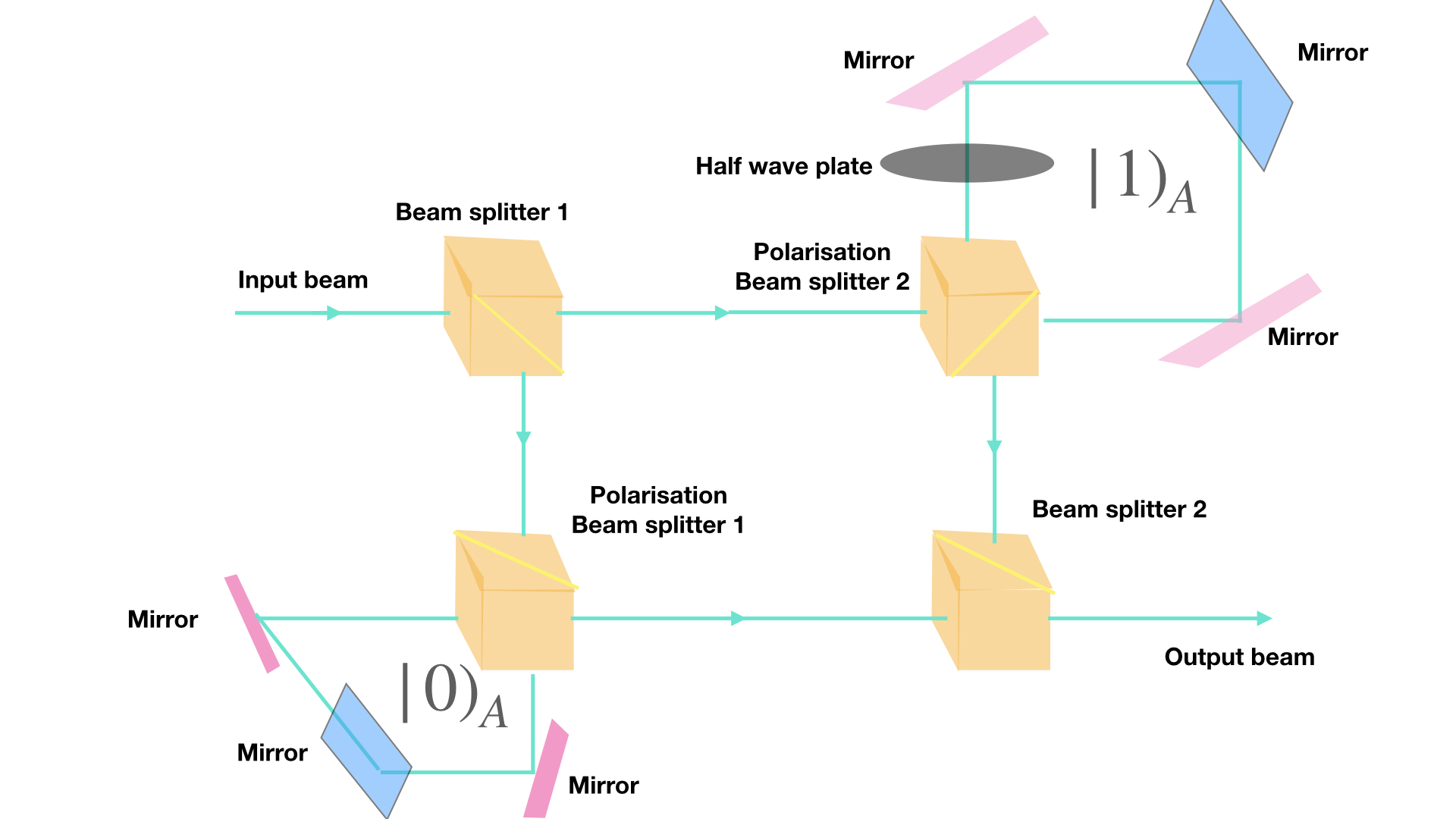}}
\caption{Schematic diagram of the setup for performing CNOT gate operation.}
\label{Sagnac_fig}
\end{figure}

In order to make a projection along the Bell beams in the polarisation and path degree of freedom, one needs to employ a CNOT gate in the two degrees of freedom. The required CNOT gate has already been experimentally implemented in \cite{Guzman16} in the experimental setup for performing classical teleportation using an entangled classical beam. To recapitulate, the CNOT gate can be implemented by using two Sagnac interferometers and a polarising beam splitter, configured as in figure (\ref{Sagnac_fig}). The path degree of freedom acts as a control--bit and polarisation degree of freedom acts as a target bit. The interferometer, corresponding to $|1)_A$ contains a half-wave plate for rotating the polarisation. Both the beams are recombined using another beam splitter. This is followed by a second beam splitter which performs the action of a Hadamard transform. The output state that corresponds to minimal post--processing is then chosen by projecting the appropriate polarisation and path, i.e., $|1)_A|V)$.
However, if per chance, the other three modes $|0)_A|H)$, $|1)_A|H)$, or $|0)_A|V)$ are measured, it also requires a corresponding transformation in the orbital angular momentum degree of freedom, which can be implemented by the programmable holographic techniques \cite{Wang17}.\\
\subsubsection{Implementation of a unitary transformation in the OAM domain by holographic techniques}
The unitary transformations $R_1^C(\pi), R_2^C(\pi), R_3^C(\pi)$ may be performed on the OAM degree of freedom of the final beam by employing programmable holographic techniques proposed in \cite{Wang17}. 

\subsection{Comparison of performance of the proposed protocol with the standard classical teleportation protocol}
\label{Comparison}
\begin{figure}[htbp]
\centerline{\includegraphics[scale=0.55]{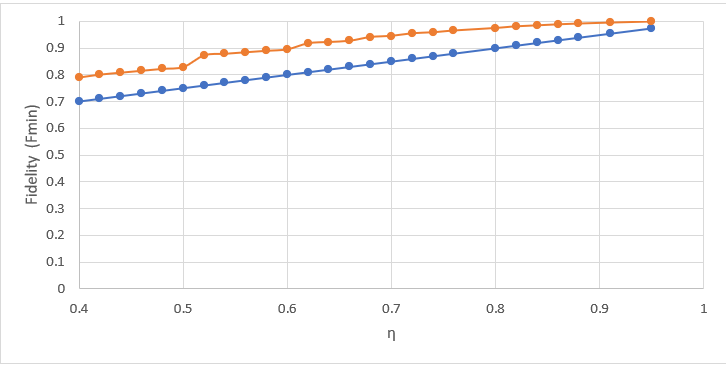}}
\caption{Orange: Fidelity $F_{\rm min}$ of the beam $\tilde{J}^C(\eta)$ with the beam $\tilde{J}^C(\eta=1)$ with respect to $\eta$ when noisy separable beam is used for teleportation. Blue: Fidelity of the two--dimensional beam $\frac{1}{2}(\mathbb{1}+\sigma^A_z)$ with $\frac{1}{2}(\mathbb{1}+\eta\sigma^C_z)$ for the same value of $\eta$. Please note that in the latter case, the shared beam used for teleportation is classically entangled (see the text for details).}
\label{Fidelity}
\end{figure}
We wish to compare the relative performance of the protocol just proposed, vis-a-vis, the standard classical teleportation protocol. Naturally, the performance depends on the {\it proximity} of the beam $\tilde{J}^C(\eta)$ ($\eta \neq 1$), to the ideal beam $\tilde{J}^C(\eta=1)$. Note that the former is the output beam in the protocol and the latter is the exact OAM equivalent of the initial beam in the path degree of freedom.
The proximity is essentially captured by {\it fidelity},  which is simply given by,
\begin{align}
    F=\Bigg\{{\rm Tr}\sqrt{\sqrt{\tilde{J}^C(\eta)}\tilde{J}^C(\eta=1)\sqrt{\tilde{J}^C(\eta)}}\Bigg\}^2.
\end{align}
The fidelity, $F$, is a function of dimensions.  For each $\eta$, we choose $d_{\rm min}$, given by eq. (\ref{dmin}). For this choice of $d_{\rm min}$, we plot the fidelity $F_{\rm min}$ as a function of $\eta$ in figure (\ref{Fidelity}). We do so because this is the worst case scenario. The fidelity would increase if we were to employ a higher dimensional beam. 

For comparison, on the same graph, we have also plotted the fidelity of $\frac{1}{2}(\mathbb{1}+\sigma^A_z)$ (in the path degree of freedom) with $\frac{1}{2}(\mathbb{1}+\eta\sigma^C_z)$, (in the two--dimensional subspace spanned by OAM modes). It is the output beam  when a mixed entangled $2\times 2$ Werner beam acts as a channel in classical teleportation.

The plots reveal that it is easier to transfer information using separable equivalent beam for the following two reasons: (i)  a SEW constitutes a more efficient channel, (ii)
     secondly, employing a SEW as a shared channel, we get higher values of fidelity for all values of $\eta$.

In summary, the advantage of the proposed protocol is that it does not require a classically entangled beam as  a channel.  In particular, for $T=1$, we have given an exact separable expansion of SEW (subsection (\ref{Separable_decomposition_1}) of section (\ref{Equivalent})). Employing this beam, a fidelity of $0\cdot 8$ is, in principle, achievable.  This concludes our discussion of application of equivalent separable beams in an information-transfer protocol.

 One might argue that  for higher $T$ values, the generation of SEW would require an increasingly large number of beams, which could be experimentally challenging. To address this issue, in the next section, we examine whether  a mixed $2\times 2$ spin--orbit coupled beam may constitute a channel for efficient information transfer. 

\section{Information transfer with mixed $2\times 2$ Werner beams }
\label{Information_noise}
We have, so far, focused on transfer of information with higher dimensional separable equivalents of $2 \times 2$ entangled Werner beams. The separable equivalents are invariably mixed. In this section, we address the question whether it is possible to transfer information efficiently by employing mixed beams in the lower dimension itself.  This question is pertinent because in most of the scenarios, \`{a} la quantum mechanics, it is implicitly assumed that the information transfer requires classically entangled light\footnote{  In most of the scenarios, state transfer from one degree of freedom to another, by employing pure classically entangled light, is aimed at transferring information from the former to the latter.}.

To address this, we consider the simplest case of $2\times 2$ Werner beams for information transfer between different degrees of freedom.   An increase in mixedness (noise) is not a major impediment since we shall see that, crucially, retrieval of the transferred information is limited entirely by the sensitivity of the detector.   

 Essentially the same protocol that was proposed in the previous section works here too except for the following modifications: (i) the $2\times 2$ noisy Werner beam acts as the channel instead of its $2\times (2T+1)$--dimensional separable equivalent beam. (ii) Secondly, information retrieval mechanism is different, depending only on the contrast in the intensity profiles. In this context, we note that noisy $d\times d$-- separable (but discordant) states have been recently used for teleportation--like protocol in the quantum regime \cite{Chen21}. The practical purpose that the protocol proposed in this section  serves is that of transferring information from one degree of freedom to another completely even when the shared channel is a noisy one.

 To describe the protocol, we choose a mixed $2\otimes 2$ Werner beam in the polarisation and OAM degrees of freedom that acts as the channel. Its coherent mode representation is given by,
 \begin{align}
 \label{Channel}
 J^{BC}(\eta)^{1/2, 1/2} \equiv \dfrac{1}{4}(\mathbb{1}-\eta\vec{\sigma}^B\cdot\vec{\sigma}^C),~\eta \in \Big[-\dfrac{1}{3},1\Big].
 \end{align}
  The superscripts $B$ and $C$ stand for polarisation and OAM respectively. The beam $J^{BC}(\eta)^{1/2, 1/2}$ is separable in the range $\eta \in \Big[-\dfrac{1}{3},\dfrac{1}{3}\Big]$.  The parameter $\eta$ is an inverse measure of noise in the channel; $\eta=0$ corresponds to the completely noisy channel, whereas $\eta=1$ corresponds to the pure channel.  The experimental techniques to generate Werner beam in the spin--orbit modes have been recently proposed in \cite{Balthazar21}.

Let $J^A=\frac{1}{2}(\mathbb{1}+\vec{\sigma}^A\cdot\vec{p})$ be the coherent mode representation of an unknown beam in the path degree of freedom, whose information content, encoded in $\vec{p}$, is to be transferred to the OAM domain. 
  The combined coherent mode representation of the three degrees of freedom is given by, 
   \begin{equation}
   \label{complete_beam_2by2}
      {J}^A\otimes J^{BC}(\eta)^{1/2, 1/2}=\frac{1}{2^3}(\mathbb{1}+\vec{\sigma}^A\cdot\vec{p})\otimes(\mathbb{1}-\eta\vec{\sigma}^B\cdot\vec{\sigma}^C).
  \end{equation}
 Since Werner beams contain a fraction of maximally entangled beam admixed with white noise, the signal--to--noise ratio of the teleported beam is given exactly by the same value $\eta$. The correlation between  polarisation and OAM degrees of freedom of (\ref{Channel}) need not be due to entanglement if $\eta \leq \frac{1}{3}$, which is the domain of interest for us. The protocol is as follows:
 
 \begin{enumerate}
     \item The experimentalist may simply project the beam given in (\ref{complete_beam_2by2}) along $|\psi_1) = \frac{1}{\sqrt{2}}\big(|0)_A|V)-|1)_A|H)\big)$. Due to this projection, the information of path degree of freedom ($A$) gets transferred to the OAM degree of freedom ($C$) and  the intensity of the beam gets reduced to one-fourth of its initial value.  This constitutes a significant simplification over quantum teleportation in that no post--processing is required. That is, the beam, after the projection, in the OAM degree of freedom is simply given by,
\begin{align}
\label{Measurement_outcome}
    \tilde{J}^C(\eta)\equiv \dfrac{1}{2}(\mathbb{1}+\eta\vec{\sigma}^C\cdot\vec{p}).
\end{align}
This situation is very much {\it unlike} to what happens in quantum mechanics which is intrinsically probabilistic.
\item Should the experimentalist choose to project along any of the other three Bell beams, as in the earlier case, the intensity will get diminished to one-fourth of its initial value as before. In addition, post-processing would be required. If the experimenter projects along one of the following three Bell beams,
  \begin{eqnarray}
&  |\psi_2)=\frac{1}{\sqrt{2}}\big\{|0)_A|H)-|1)_A|V)\big\};  \nonumber\\ &|\psi_3)=\frac{1}{\sqrt{2}}\big\{|0)_A|H)+|1)_A|V)\big\};\nonumber\\
&|\psi_4)=\frac{1}{\sqrt{2}}\big\{|0)_A|V)+|1)_A|H)\big\},
  \end{eqnarray}
 on the path and polarisation degree of freedom, the transformations to be applied on the OAM degree of freedom are $ \sigma_{1}^C, \sigma_{2}^C, \sigma_{3}^C$, in the same order.
 \end{enumerate}
  This brings out the fundamental difference between quantum and classical teleportations. The measurement in the former is probabilistic, whereas, in the latter, is deterministic. As a consequence, while the post-processing (application of unitary operations) is inevitable in the former case, in the latter, the projection along $|\psi_1)$ does away with any post-processing.
 
The projection along  Bell beams can be performed in the path and polarisation degrees of freedom by employing Sagnac interferometers and a half wave plate, which performs a CNOT gate operation, It has been discussed in detail in \cite{Guzman16} and already recapitulated in the subsection (\ref{Sagnac}) of the section (\ref{Remote state preparation unknown qubit}).   If necessary, the unitary transformations on the OAM modes, corresponding to the projections along $|\psi_2), |\psi_3)$ and $|\psi_4)$ can be performed using holographic techniques proposed in \cite{Wang17}.

 We have, so far, discussed the transfer of  information  from the path to the OAM degree of freedom. We now address how to retrieve this information in the next subsection. 
 \subsection{Retrieval of information in the OAM degree of freedom}
We observe from the equation (\ref{Measurement_outcome}) that the effect of noise (i.e., $\eta$) is reflected in rescaling of $\vec{p}$ to $\eta\vec{p}$ ($|\eta| <1$), when the information is transferred from the path to the OAM degree of freedom. Experimentally, the value of $\eta$ manifests as signal-to--noise ratio. Thanks to the experimental advances, this is not a major disadvantage since the process of information retrieval has become considerably simpler.  Speaking algebraically, in order to retrieve the information, a measurement of $\vec{\sigma}^C\cdot\hat{n}$ is to be performed, followed by a rescaling with respect to $\eta$ (the case $\eta =0$ is a singularity from which no information retrieval is possible). This would offset the diminution due to noise. Thus,  the measurement of a corresponding observable  $\frac{1}{\eta}\vec{\sigma}^C\cdot\hat{n}$, in principle, may be employed to extract $\vec{p}$, which encodes the information. The measurement of $\frac{1}{\eta}\vec{\sigma}^C\cdot\hat{n}$ is limited exclusively by the resolution of the detector.

\begin{figure}[htbp]
\centerline{\includegraphics[scale=0.14]{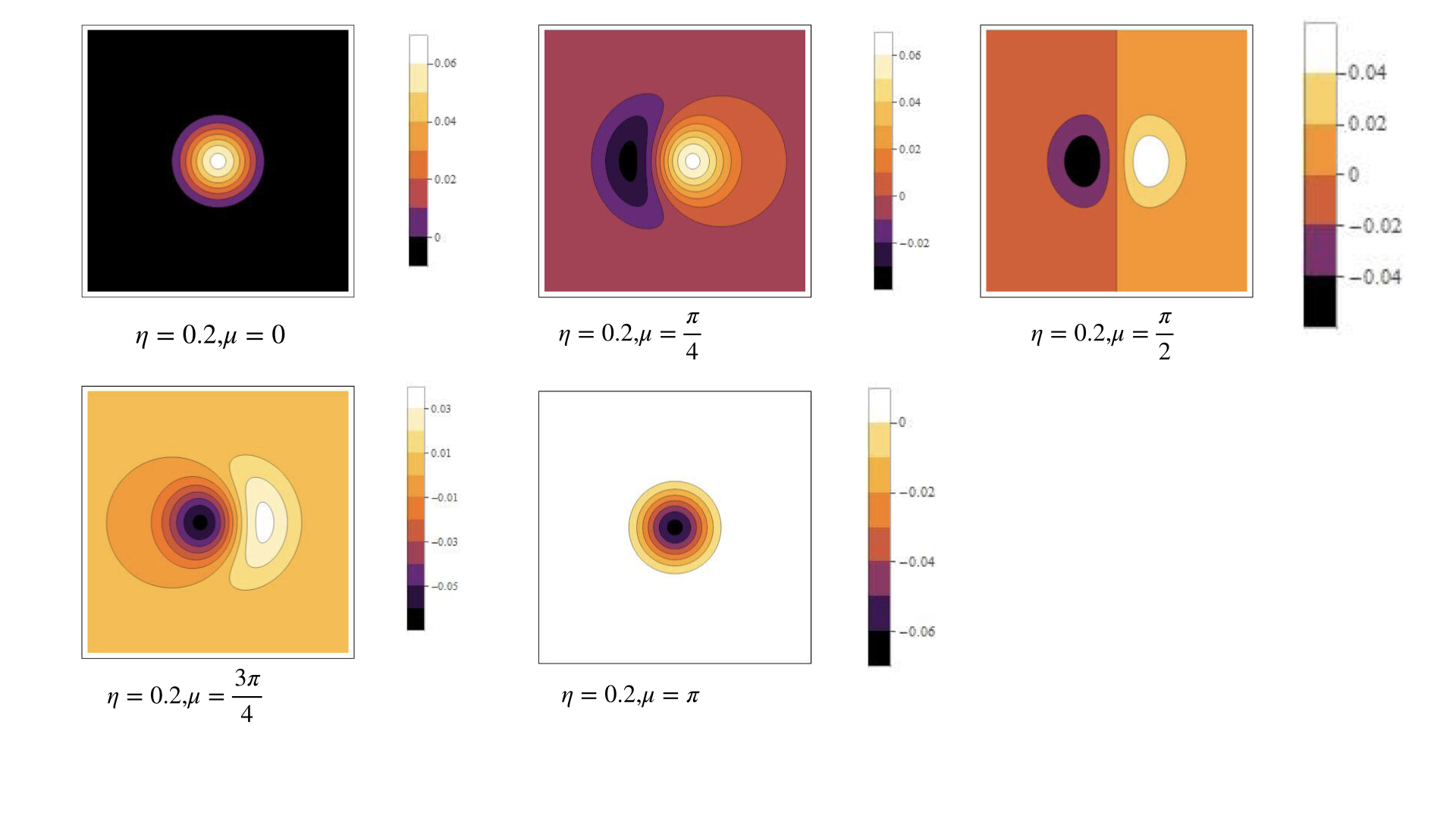}}
\caption{Plot of $I_{\rm diff}(r, \phi)$ in the transverse plane for $\eta=0.2$ and different  values of $\mu$. Note that the intensity profile changes with $\mu$. The scale of the plot is of the order of 0.06.}
\label{alpha=0.2}
\end{figure}

\begin{figure}[htbp]
\centerline{\includegraphics[scale=0.15]{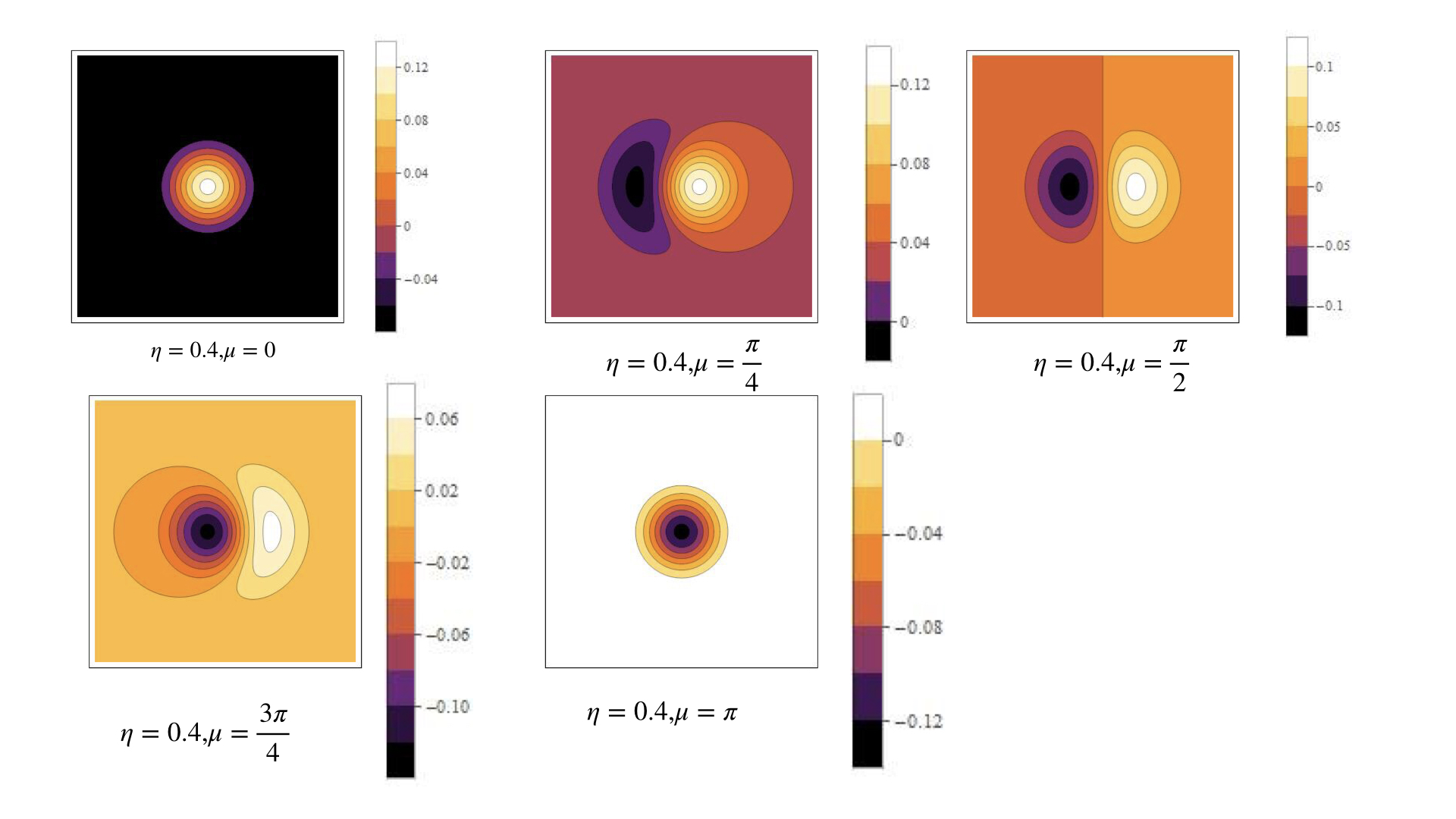}}
\caption{Plot of $I_{\rm diff}(r, \phi)$ in the transverse plane for $\eta=0.4$ and different  values of $\mu$. The intensity profile changes with $\mu$. The scale of the plot is of the order of 0.12.}
\label{alpha=0.4}
\end{figure}

\begin{figure}[htbp]
\centerline{\includegraphics[scale=0.15]{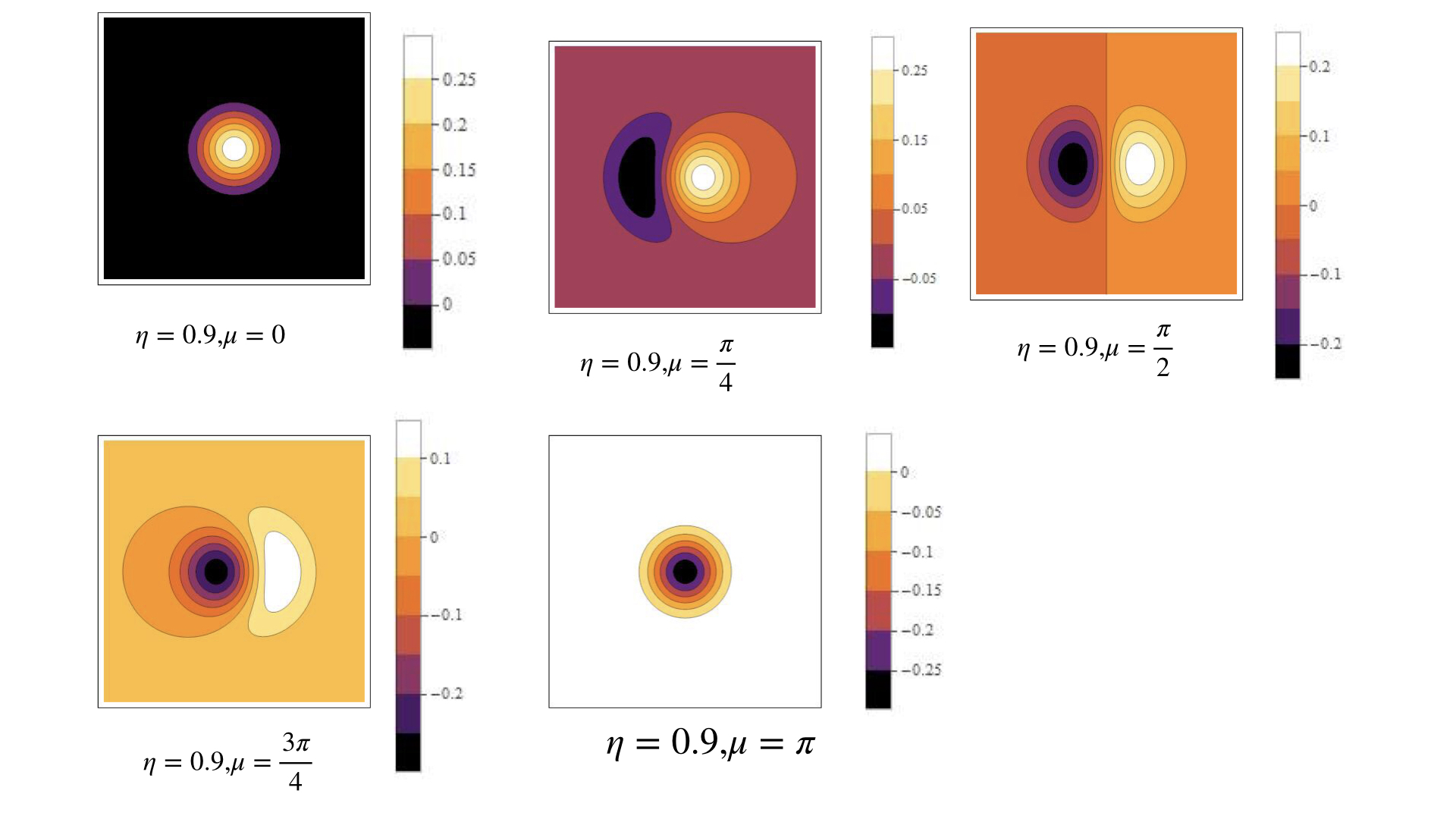}}
\caption{Plot of $I_{\rm diff}(r, \phi)$ in the transverse plane for $\eta=0.9$ and different  values of $\mu$. The intensity profile changes with $\mu$. The scale of the plot is of the order of 0.25.}
\label{alpha=0.9}
\end{figure}

We work out an explicit example. A natural measure of information in a light beam is the variation in the pattern against a background corresponding to the completely mixed channel. Let the two-dimensional OAM subspace be spanned by the modes $|{\rm LG}_{00})$ and $|{\rm LG}_{01})$ and $\vec{p}=(\sin\mu, 0, \cos\mu)^T$. Then, the coherent mode representation of the beam in two--dimensional subspace of OAM degree of freedom is,
\begin{align}
\label{SNR_beam}
\dfrac{1}{2}(\mathbb{1}+\eta\vec{\sigma}_3\cdot\hat{p})\equiv\dfrac{1+\eta}{2}|\psi_1)(\psi_1|+\dfrac{1-\eta}{2}|\psi_2)(\psi_2|,
\end{align}
where the beams $|\psi_1)$ and $|\psi_2)$ are given by,
\begin{align}
|\psi_1) &=\cos\dfrac{\mu}{2}|{\rm LG}_{00})+\sin\dfrac{\mu}{2}|{\rm LG}_{01}),\nonumber\\
|\psi_2) &=-\sin\dfrac{\mu}{2}|{\rm LG}_{00})+\cos\dfrac{\mu}{2}|{\rm LG}_{01}).
\end{align}
For a given value of $\eta$, let $I_{\eta}(r, \phi)$ represent the transverse intensity of the beam given in equation (\ref{SNR_beam}). Obviously, $\eta =0$ corresponds to the completely mixed channel. In order to study the  effect of $\eta$ on signal-to-noise ratio of the beam represented by (\ref{SNR_beam}), we plot the difference in intensities $I_{\rm diff}(r, \phi) \equiv I_{\eta}(r, \phi)-I_0(r, \phi)$, which is given by,  
\begin{align}
I_{\rm diff}(r, \phi) \equiv& \frac{1+\eta}{2}\Big|\cos\frac{\mu}{2}{\rm LG}_{00}(r, \phi)+\sin\frac{\mu}{2}{\rm LG}_{01}(r, \phi)\Big|^2\nonumber\\
+&\frac{1-\eta}{2}\Big|-\sin\frac{\mu}{2}{\rm LG}_{00}(r, \phi)+\cos\frac{\mu}{2}|{\rm LG}_{01}(r, \phi)\Big|^2\nonumber\\
-&\frac{1}{2}\Big(|{\rm LG}_{00}(r, \phi)|^2+{\rm LG}_{01}(r,  \phi)|^2\Big).
\end{align}
 We plot $I_{\rm diff}(r, \phi)$ for three values of $\eta$, {\it viz.}, $0.2$, $0.4$ and $0.9$. Note that only $\eta=0.2$ corresponds to a separable beam. For each value of $\eta$, we have plotted $I_{\rm diff}(r, \phi)$ for $\mu=0, \frac{\pi}{4}, \frac{\pi}{2}, \frac{3\pi}{4}$ and $\pi$. 
The following two features are of significance to us:
\begin{enumerate}
\item {\bf Effect of variation of $\eta$ for a fixed $\mu$:}\\
A comparison of figures (\ref{alpha=0.2}), (\ref{alpha=0.4}) and (\ref{alpha=0.9}) shows that the contrast in the image increases with increase in $\eta$. We observe that the shape of the intensity-profile remains unchanged with increase in $\eta$, with the contrast being increasingly monotonic with purity of the beam. That is, the only change is in the scale of the plot. This substantiates our point that the same information can be retrieved from noisy beams provided a detector with sufficient resolution is available.
\item {\bf Effect of variation of $\mu$ for a fixed $\eta$:}\\
For each $\mu$, there is a characteristic pattern of $I_{\rm diff}(r, \phi)$, as may be seen from figures (\ref{alpha=0.2}), (\ref{alpha=0.4}) and (\ref{alpha=0.9}). The change in the pattern with respect to $\mu$ contains information encoded in different linear combinations of ${\rm LG}_{01}(r, \phi)$ and ${\rm LG}_{0-1}(r, \phi)$. 
\end{enumerate}

We may now conclude the section by noting that a variation in $\eta$ changes the contrast of the pattern, whereas a variation in $\mu$ changes the profile of the pattern.   We now turn our attention to a more conventional application, {\it viz.}, variational quantum algorithms.
 \section{Variational quantum  classifier algorithms}
\label{VQA}
 Recently, variational quantum algorithms have attracted a lot of attention \cite{wan2017quantum, farhi2018classification, rebentrost2018quantum, schuld2019quantum, havlivcek2019supervised}, as they provide speedups as compared  to their classical counterparts. The protocols consist of two broad varieties: (I) those employing multi-qubit systems, and, (II) those  employing a single qu$N$it system ($N$-- dimensional quantum states).  The former protocols employ entangling gates, e.g., CNOT gate, Toffolli gate, etc.  The latter  protocols, however,  do not employ entangling gates. Although the two are physically distinct, mathematically they are essentially equivalent.  All the gates acting on $N$-qubit systems map to some operations belonging to the special unitary group of dimension $D(\equiv 2^N)$ for a qudit system of dimension $D$. 

We first describe in brief the variational quantum classifier algorithm, proposed in \cite{Adhikary20a} for a qu$N$it system, for the purpose of illustration. For a detailed discussion of this and associated protocols, we refer the reader to \cite{wan2017quantum, farhi2018classification, rebentrost2018quantum, schuld2019quantum, havlivcek2019supervised, Adhikary20a}.

 Consider a data set $\mathbb{S} = \{ ({\bf x}, f({\bf x}))\}$. Each element of $\mathbb{S}$ is an ordered pair $({\bf x}, f({\bf x}))$, where ${\bf x}$ is an input vector  of arbitrary dimension  and $f({\bf x})$ is its associated label indicating the particular class the input belongs to. Thus, $f$ is a mapping of each input vector to a label, belonging to a set: $\mathbb{L} = \{ l_1, l_2, \cdots, l_N\}$; $f: {\bf x} \rightarrow \mathbb{L}$. Each label corresponds to an output class, whose total number is $N$.  Supervised learning aims at training a machine using a subset $\mathbb{T}$ chosen from the given dataset such that the machine can infer correct labels for the train set $\mathbb{T}$ as well as the test set $\mathbb{S}-\mathbb{T}$. That is, we require the machine to return a function $f^*$ so that $f^*({\bf x}) = f({\bf x})$ for maximum number of input vectors. 

A typical quantum protocol for supervised learning, that employs variational algorithms, can be divided into the following three stages:
\subsection{State preparation in the quantum domain}
\label{sp}
To initialise the protocol, let there be a single qu$N$it  belonging to the $N$-- dimensional Hilbert space $\mathcal{H}^N$. It is ensured that the dimension of the Hilbert space is equal to the cardinality of $\mathbb{L}$, i.e., the total number of output classes. This is essential for the scheme proposed in \cite{Adhikary20a}. The input vectors, ${\bf x}$, will be represented in $\mathcal{H}^N$ by $\ket{\psi(\bf x)}$. The encoding scheme employed to obtain $\ket{\psi(\bf x)}$ is given as,

   \begin{align}
   \label{encoding 1}
 \ket{\psi(\bf x)} &= e^{i \overline{S}_3 (\sum_{j=1}^d w_j x_j)} H^{(N)} \ket{0} \\ \nonumber
   &\equiv Z (w_1, w_2, \cdots, w_d) H^{(N)} \ket{0}
   \end{align}

Here, $x_j$ represents the $j$-th component of the input vector ${\bf x}$, $\overline{S}_3$ is the diagonal matrix $diag (-(N-1)/2, \cdots,(N-1)/2)$, $\ket{0} = (1,0,0, \cdots, 0)^\dag$ and $H^{(N)}$ is the generalized Hadamard gate.

  The variables $\{w_i; i=1,2, \cdots, d\}$ are free parameters that can be tuned, during training. It is, thus, a parameter dependent state preparation procedure.
  \subsection{Parametrized unitary operations in the quantum domain}
\label{puo}

 Input vectors ${\bf x}$ are encoded in states of a single qu$N$it. The most general unitary operation that can be applied to such a state belongs to the $SU(N)$ group. Since the group $SU(N)$ has $(N^2-1)$ generators, the most general unitary transformation belonging to it admits a generalised Euler angle parametrisation as \cite{Byrd_1998}:
\begin{equation}
   \label{group}
    U(N)(\{\alpha_j\}) = e^{i \lambda_{1} \alpha_{1}} e^{i \lambda_2 \alpha_{2}}\cdots e^{i \lambda_{(N^2 -1)} \alpha_{(N^2 -1)}}.
\end{equation}
 The set $\{ \alpha_j; j = 1, 2, \cdots, (N^2-1)\}$ constitutes a set of learnable parameters while the set $\{ \lambda_j\}$ consists of the standard generators of $SU(N)$.

\subsection{Measurement and decision functions in the quantum domain}
\label{measurement}
The final stage of the algorithm involves prediction of the label for an input vector by the classifier. In \cite{Adhikary20a}, this decision making is determined by a projective measurement on the transformed state $ \ket{\tilde{\psi}(\bf x)} = U(N)\ket{\psi(\bf x)}$ in the basis $\{|0\rangle, \cdots, |N-1\rangle\}$. Thus, measurement of any non-degenerate operator, say, $A= \sum_{a=0}^{N-1}a|a\rangle\langle a|$  in this basis may be employed. 
  
In order to predict the label associated with ${\bf x}$, the complete set of probabilities $\{ p_a={\rm Tr} \big(\pi_a \ket{\tilde{\psi}(\bf x)}\big)\}$, corresponding to all outcomes, for the state $\ket{\tilde{\psi}(\bf x)}$ would have to be constructed. The following rule is stipulated:

\begin{equation}
\label{Rule}
    p_b = max\{ p_a \} \rightarrow f^*({\bf x}) = l_b,
\end{equation}
i.e., the predicted label will be $l_b$ iff $S_3 = b$ is the most likely outcome of the measurement.     The predicted label is not necessarily the same as actual label. Hence to minimize mismatches, one needs to train the circuit. A sample ${\bf x}$ is said to be correctly classified iff $f^*({\bf x}) = f({\bf x})$.

\begin{figure}[htbp]
\centerline{\includegraphics[scale=0.21]{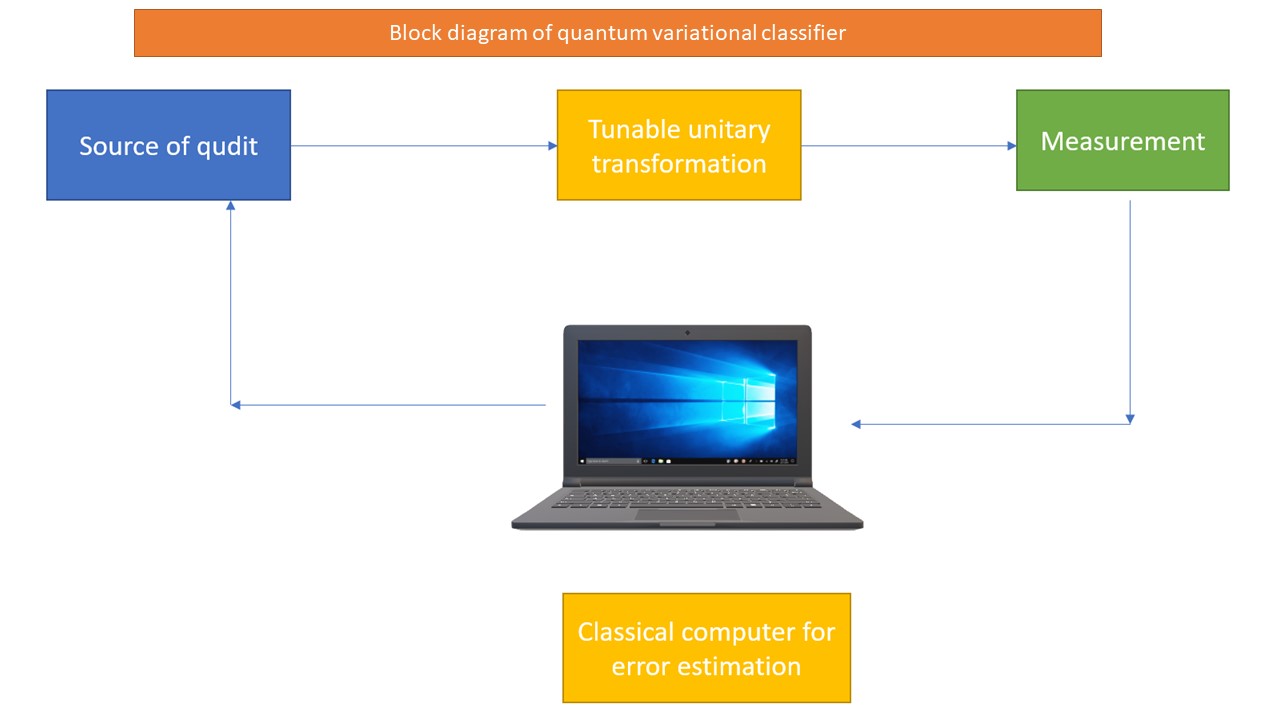}}
\caption{Schematic diagram of the setup of a quantum variational classifier.  Parametrised unitary transformations act on a qudit, followed by measurements in the measurement basis. Error estimation is done by a classical computer, which is used as a feedback to the source generating the qudit.}
\label{Quantum variational classifier}
\end{figure}

This concludes the forward part of the computation. After this, an error estimation is done with a classical computer. The training process aims at minimizing the error function ${\bf E}$. It is carried out on a classical computer since ${\bf E}$ is a classical function.  After each epoch, the training parameters $\{ w_i; i=1, 2, \cdots, d\}$ and $\{\alpha_i; i=1, 2, \cdots, (N^2-1)\}$ are updated to minimize ${\bf E}$, say, by using the classical gradient descent technique. This is the feedback part of the computation. After that, the forward part of the computation is carried out once again. This process is repeated until the value of ${\bf E}$ converges to a minimum. At that stage the values of the free parameters are frozen and training is complete. The outline of the setup of variational quantum classifier is shown in figure (\ref{Quantum variational classifier}).

\section{Implementation of variational quantum classifier algorithm with classical light}
\label{meth}
It is evident that due to robust generation, manipulation and detection techniques of OAM modes, quantum algorithms that require only parallelism can be simulated with classical light efficiently. In particular, we propose, in this section,  how variational quantum algorithms can be implemented with OAM modes of classical light, through the example of the quantum classifier circuit discussed in the preceding section.

As before, consider the dataset $\mathbb{S} = \{ {\bf x}, f({\bf x}) \}$; ${\bf x} \in \mathbb{R}^d$, $f: {\bf x} \rightarrow \mathbb{L} = \{ l_1, l_2, \cdots, l_N\}$. $\mathbb{L}$ is the set of labels. In what follows, we propose the setup for implementation of quantum classifier algorithm stage-by-stage by employing OAM modes:

\subsection{Beam preparation with classical light} 
 First of all, we replace $|\psi(x)\rangle$ (give in equation (\ref{encoding 1})) by $|\psi({\bf x}))$, which is obtained by superposition of  $N$ orthonormal Laguerre Gauss modes of classical light.  The beam $|\psi({\bf x}))$ can be directly prepared  using spatial light modulator \cite{Forbes16}. That is to say,
\begin{eqnarray}
|\psi ({\bf x})\rangle = \sum_i\mu_i|\psi_i\rangle\mapsto  \sum_i\mu_i|{\rm LG}_{0i}) \equiv |\psi({\bf x})).
\end{eqnarray}
There is no need to first prepare the beam corresponding to  $|0\rangle $ (i.e., $|LG_{00}) $ beam, and follow it up by some unitary transformations to arrive at $|\psi({\bf x}))$.

\subsection{Parametrized unitary operations with classical light} 
The tunable unitary transformations on the OAM modes can be performed by using holographic techniques proposed in \cite{Wang17}. This method allows, in principle, the implementation of any unitary transformation on the OAM modes. To describe briefly, the input beam, which is already in a superposition of OAM modes, passes through a hologram, designed to perform the desired transformation. Thereafter, a second hologram reduces the cross-talk between different modes. If there are $N$ modes involved, the lower bound on efficiency of this technique is found to be $\frac{1}{N^{3/2}}$ \cite{Wang17}. 

\subsection{Measurement and decision functions with classical light} 
We shall now show how the task of measurements and determination of decision functions can be accomplished with OAM modes. First, we perform a log-polar transform on the OAM beam, emergent from the tunable hologram, so that intensity, corresponding to each Laguerre-Gauss mode, is distributed  in the transverse direction. It can be captured via a charge coupled device (CCD). The normalised intensity is formally analogous to the quantum probability. Hence, equation (\ref{Rule}) translates to the fact that the mode corresponding to $b$ (in equation (\ref{Rule})) has the greatest intensity.  The output of that CCD is given as a feedback to the computer that creates the pattern for spatial light modulator.

\begin{figure}[htbp]
\centerline{\includegraphics[scale=0.21]{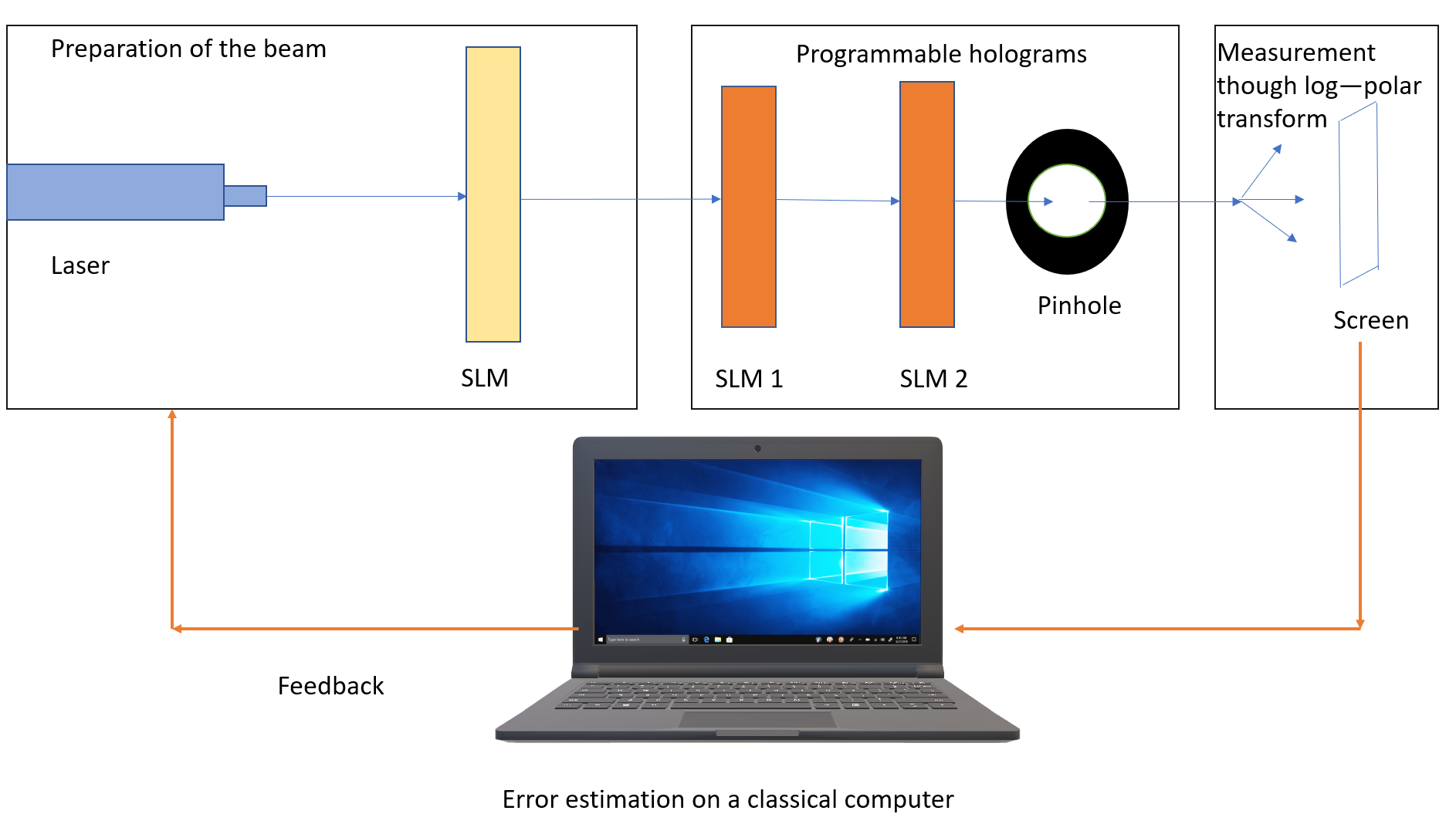}}
\caption{Schematic diagram of the setup of a classical variational classifier. An OAM beam can be prepared by using  spatial light modulator (SLM), which passes through programmable holograms that implement parametrised unitary transformation. Finally, intensity measurement in different modes can be performed using log--polar transformation. Error estimation is done by a classical computer, which is used as a feedback to the SLM generating the OAM beam.}
\label{Classical variational classifier}
\end{figure}

The backward part of the computation is performed on a classical computer whether the protocol is implemented on a classical or a quantum platform. So, it requires no additional setup in this case. The same steps, as discussed in the previous section, would do the job in this case as well.  The outline of a setup for variational classifier using OAM modes is shown in figure (\ref{Classical variational classifier}).
We have, so far, shown a setup for the protocol, originally proposed for a qu$N$it system. However, many variational algorithms have been proposed for $N$-qubit systems. They can also be implemented with OAM modes of classical light with an equal ease. Obviously, the composite $N$--qubit system in these algorithms belongs to a $2^N$--dimensional Hilbert space where all the gate operations are done. Thanks to isomorphism of vector spaces of identical dimensions, the corresponding gates can be identified for a single qudit system of the same dimension $2^N$. 
 
 As an example, if we consider a two-qubit system, the most general transformation acting on it belongs to $SU(4)$. A natural choice of basis for the operators is as follows:
\begin{align}
\{\mathbb{1}, \sigma_{\mu}^A\otimes \mathbb{1}, \mathbb{1}\otimes\sigma_{\nu}^B, \sigma_{\mu}^A\otimes\sigma_{\nu}^B\}; \mu, \nu \in \{1, 2, 3\}.
\end{align}
 One may first make the following mapping of the basis elements from the Hilbert space of a two--qubit to the Hilbert space of a single four-level system:
\begin{align}
|i_1i_0\rangle \mapsto \big|2i_1+i_0\big\rangle;~i_0, i_1\in \{0, 1\}.
\end{align}
With this mapping, all the states, observables and transformation matrices get mapped from the two-qubit systems to a single four-level system.   This procedure admits straightforward generalisation to $N$--qubit systems.

Thus, the implementation of these protocols with OAM modes is a two--step process. First, the basis vectors of tensor product of $N$ two--dimensional Hilbert spaces should be mapped to those of the $2^N$--dimensional Hilbert space of a single qudit. Secondly, the basis vectors of $2^N$--dimensional Hilbert space can be mapped to $2^N$ OAM modes of classical light.
 
  This example shows that there is a class of machine learning algorithms for different tasks that can be implemented by employing OAM modes of classical light.  We have discussed but one particular example. Of course, implementations of tunable unitary transformations become challenging with increase in dimensions of Hilbert space (i.e., the number of modes involved). The modes have to be chosen in such a way that their patterns can be resolved after log--polar transformation. Nevertheless, the need for both-- a single photon source and a single photon detector, that are quintessential for implementation of the protocols on a quantum platform--is eliminated.  Additionally, the simulability of variational algorithms with OAM modes provides a motivation for engineering of such devices that can robustly perform tunable transformations.



\section{Conclusion}
\label{Conclusion}
    In conclusion, we have put forth a new class of beams, termed as {\it equivalent beams}. Employing these beams, we have proposed a protocol for transfer of information from one degree of freedom of a classical beam to another without the need for classical entanglement. Production of equivalent beams is not experimentally difficult \cite{tuan2018realizing}. The advantage of the protocol is that the information processing can be done with highly mixed beams as one employs higher dimensionality. We have next considered variational quantum algorithms, and  outlined a proof--of--principle setup for their implementation with OAM modes of light. 

This work shows that there are many so--called quantum phenomena that can be simulated with classical light. With the availability of robust generation, manipulation and detection techniques,  OAM modes offer versatile platform to perform these tasks, at times even without classical entanglement. The setups for many other quantum phenomena and algorithms will be proposed in a separate paper.
\section*{Acknowledgement}
We thank the referee for several insightful comments and a careful critique of our work. We also thank the referee for bringing several relevant works to our attention. We thank Rajni Bala for several insightful discussions and helpful suggestions.   Sooryansh thanks  the Council for Scientific and Industrial Research (Grant no. -09/086 (1278)/2017-EMR-I) for funding his research.
 \section*{Data availability}
No data were generated or analysed in the present research.
\section*{Disclosures}
 The authors declare no conflicts of interest.


\begin{appendix}
\section{Generation of SEW}
\label{Separable_appendix}
In this appendix, in order to generate SEW, we present two sets of incoherent superpositions of pure separable beams. The first one,  corresponding to twelve beams, has the following coherent mode representation:
\begin{align}
    J^{(2)}_c\equiv \frac{1}{12}\sum_{(ij)}& \Bigg\{\Big|\frac{1}{\sqrt{2}}(\sigma_i^A+\sigma_j^A)=+1\Big)\Big(\frac{1}{\sqrt{2}}(T_i^B+T_j^B)=-T\Big|\nonumber\\
    +&\Big|\frac{1}{\sqrt{2}}(\sigma_i^A+\sigma_j^A)=-1\Big)\Big(\frac{1}{\sqrt{2}}(T_i^B+T_j^B)=+T\Big|\nonumber\\
    +& \Big|\frac{1}{\sqrt{2}}(\sigma_i^A-\sigma_j^A)=+1\Big)\Big(\frac{1}{\sqrt{2}}(T_i^B-T_j^B)=-T\Big|\nonumber\\
    \label{twelve beams}
    +&\Big|\frac{1}{\sqrt{2}}(\sigma_i^A-\sigma_j^A)=-1\Big)\Big(\frac{1}{\sqrt{2}}(T_i^B-T_j^B)=+T\Big|\Bigg\},
\end{align}
where $(i, j)$ are distinct and vary cyclically over $(1, 2, 3)$. 
The second set, involving eighteen beams, is given by:
\begin{align}
\label{eighteen_beams}
    J^{(3)}_c\equiv \dfrac{1}{3}J^{(1)}_c+\dfrac{2}{3}J^{(2)}_c,
\end{align}
where $J^{(1)}_c$ has been defined in equation (\ref{Six_beams}). The second set yields us a beam, which is quite close to the desired equivalent beam. This has been captured by the norm of their difference, already been shown in table (\ref{Proximity}).

\section{Information--transfer protocol from one degree of freedom to another}
\label{Algebra}
In this appendix, we work out the algebra of the operators underlying the protocol, proposed in section (\ref{Remote state preparation unknown qubit}). We start with the initial beam, given in equation (\ref{Complete_beam_2by2T+1}),
\begin{align}
    J^A\otimes J^{BC}(\eta)^{1/2, T}\equiv \dfrac{1}{2^2(2T+1)}(\mathbb{1}+\vec{\sigma}^A\cdot\vec{p})\otimes (\mathbb{1}-\eta\vec{\sigma}^B\cdot\hat{T}^C).\nonumber
\end{align}

A measurement along the beam $\frac{1}{\sqrt{2}}\big(|0)_A|V)-|1)_A|H)\big)$ corresponds to projecting the initial beam onto the operator,
\begin{align}
\label{BellProj1}
 \Pi_1\equiv   \frac{1}{2}\big\{|0)_A|V)-|1)_A|H)\big\}\big\{{}_A(0|(V|-{}_A(1|(H|\big\}.
\end{align}
The projection operator $\Pi_1$ can be expressed in terms of Pauli operators as follows:
\begin{align}
 |0)_A(1| &\equiv \dfrac{1}{2}(\sigma_1^A+i\sigma_2^A);~~ |1)_A(0| \equiv \dfrac{1}{2}(\sigma_1^A-i\sigma_2^A);\nonumber\\
 \label{PathPauli}
    |0)_A(0|&\equiv \dfrac{1}{2}(\mathbb{1}+\sigma_3^A);~~ |1)_A(1|\equiv \dfrac{1}{2}(\mathbb{1}-\sigma_3^A),
\end{align}
and,
\begin{align}
|H)(V| &\equiv \dfrac{1}{2}(\sigma_1^B+i\sigma_2^B);~~ |V)(H| \equiv \dfrac{1}{2}(\sigma_1^B-i\sigma_2^B);\nonumber\\
\label{PolPauli}
    |H)(H|&\equiv \dfrac{1}{2}(\mathbb{1}+\sigma_3^B);~~ |V)(V|\equiv \dfrac{1}{2}(\mathbb{1}-\sigma_3^B),
\end{align}
whence,
\begin{align}
    \Pi_1\equiv \dfrac{1}{4}(\mathbb{1}-{\sigma}_1^A{\sigma}_1^B-{\sigma}_2^A{\sigma}_2^B-{\sigma}_3^A{\sigma}_3^B)=\dfrac{1}{4}(\mathbb{1}-\vec{\sigma}^A\cdot\vec{\sigma}^B).
\end{align}
The final beam, in the OAM degree of freedom, is then simply given by,
\begin{align}
    &{\rm Tr}_{AB}\Big\{\Pi_1 \Big(J^A\otimes J^{BC}(\eta)^{\frac{1}{2}, T}\Big)\Big\} \equiv \nonumber\\
   = &{\rm Tr}_{AB}\Big\{\dfrac{1}{4}(\mathbb{1}-\vec{\sigma}^A\cdot\vec{\sigma}^B)\dfrac{1}{2^2(2T+1)}(\mathbb{1}+\vec{\sigma}^A\cdot\vec{p})\otimes (\mathbb{1}-\eta\vec{\sigma}^B\cdot\hat{T}^C)\Big\}\nonumber\\
     \label{Calculation}
     =&\dfrac{1}{4}\times \dfrac{1}{2T+1}(\mathbb{1}+\eta\hat{T}^C\cdot\vec{p}).
\end{align}
The factor of $\frac{1}{4}$ represents the reduction in intensity by a factor of 4. Similarly, the final beams for the other three projections can be found out in a similar manner. 
\begin{enumerate}
    \item {\bf Projection along $|\psi_2)\equiv\dfrac{1}{\sqrt{2}}\big\{|0)_A|H)-|1)_A|V)\big\}$}\\
    \noindent As before, by employing the relations given in equations (\ref{PathPauli}) and (\ref{PolPauli}), the corresponding projection operator $\Pi_2=|\psi_2)(\psi_2|$ can be written in the form of Pauli operators as, 
    \begin{align}
      \Pi_2 \equiv \dfrac{1}{4}(\mathbb{1}-\sigma^A_1\sigma_1^B+\sigma_2^A\sigma_2^B+\sigma_3^A\sigma_3^B). 
    \end{align}
    The final beam, in the OAM degree of freedom, after the projection $\Pi_2$, can be found by evaluating the following expression,
\begin{align}
    &{\rm Tr}_{AB}\Big\{\Pi_2 \Big(J^A\otimes J^{BC}(\eta)^{\frac{1}{2}, T}\Big)\Big\} \equiv \nonumber\\
    \label{Calculation2}
    =&\dfrac{1}{4}\times\dfrac{1}{(2T+1)}\Big\{\mathbb{1}+\frac{\eta}{T}(T_1p_1-T_2p_2-T_3p_3)\Big\}.
\end{align}

This is followed by a unitary transformation $e^{iT_1^C\pi}$ in the OAM domain.The transformation properties are as follows,
\begin{align}
    e^{iT^C_i\pi}T^C_ie^{-iT^C_i\pi} &=T^C_i\nonumber\\
        e^{iT^C_i\pi}T^C_je^{-iT^C_i\pi} &=-T^C_j\nonumber\\
            e^{iT^C_i\pi}T^C_ke^{-iT^C_i\pi} &=-T^C_k,
\end{align}
where $(i, j, k)$ vary cyclically over $(1, 2, 3)$. The beam, after the transformation, becomes $\dfrac{1}{(2T+1)}(\mathbb{1}+\eta\hat{T}^C\cdot\vec{p})$.
 the beam after the transformation becomes $\dfrac{1}{(2T+1)}(\mathbb{1}+\eta\hat{T}^C\cdot\vec{p})$.
 
    \item {\bf Projection along $|\psi_3)\equiv\dfrac{1}{\sqrt{2}}\big\{|0)_A|H)+|1)_A|V)\big\}$}\\
  \noindent  As before, by employing the relations given in equations (\ref{PathPauli}) and (\ref{PolPauli}), the corresponding projection operator $\Pi_3=|\psi_3)(\psi_3|$ can be written in the form of Pauli operators as, 
  \begin{align}
       \Pi_3 \equiv \dfrac{1}{4}(\mathbb{1}+\sigma^A_1\sigma_1^B-\sigma_2^A\sigma_2^B+\sigma_3^A\sigma_3^B).
  \end{align}
     The final beam, in the OAM degree of freedom, after the projection $\Pi_3$, can be found by evaluating the following expression,
\begin{align}
    &{\rm Tr}_{AB}\Big\{\Pi_3 \Big(J^A\otimes J^{BC}(\eta)^{\frac{1}{2}, T}\Big)\Big\} \equiv \nonumber\\
    \label{Calculation3}
    =&\dfrac{1}{4}\times\dfrac{1}{(2T+1)}\Big\{\mathbb{1}+\frac{\eta}{T}(-T_1p_1+T_2p_2-T_3p_3)\Big\}.
\end{align}

This is followed by a unitary transformation $e^{iT_2^C\pi}$ in the OAM domain. After the transformation, the beam becomes $\frac{1}{2T+1}(\mathbb{1}+\hat{T}^C\cdot\hat{p})$.
    \item {\bf Projection along $|\psi_4)\equiv\dfrac{1}{\sqrt{2}}\big\{|0)_A|V)+|1)_A|H)\big\}$}\\
  \noindent  As before, by employing the relations given in equations (\ref{PathPauli}) and (\ref{PolPauli}), the corresponding projection operator $\Pi_4=|\psi_4)(\psi_4|$ can be written in the form of Pauli operators as, 
  \begin{align}
       \Pi_4 \equiv \dfrac{1}{4}(\mathbb{1}+\sigma^A_1\sigma_1^B+\sigma_2^A\sigma_2^B-\sigma_3^A\sigma_3^B).
  \end{align}
     The final beam, in the OAM degree of freedom, after the projection $\Pi_4$, can be found by evaluating the following expression,
\begin{align}
    &{\rm Tr}_{AB}\Big\{\Pi_4 \Big(J^A\otimes J^{BC}(\eta)^{\frac{1}{2}, T}\Big)\Big\} \equiv \nonumber\\
    \label{Calculation2}
    =&\dfrac{1}{4}\times\dfrac{1}{(2T+1)}\Big\{\mathbb{1}+\frac{\eta}{T}(-T_1p_1-T_2p_2+T_3p_3)\Big\}.
\end{align}
This is followed by a unitary transformation $e^{iT_3^C\pi}$ in the OAM domain. 
\end{enumerate}
\end{appendix}

\begin{thebibliography}{10}
\newcommand{\enquote}[1]{``#1''}

\bibitem{Deutsch85}
D.~Deutsch, \enquote{Quantum theory, the church-turing principle and the
  universal quantum computer,} {\protect\JournalTitle{Proceedings of the Royal
  Society of London. Series A, Mathematical and Physical Sciences}}
  \textbf{400}, 97 (1985).

\bibitem{Deutsch92}
D.~Deutsch and R.~Jozsa, \enquote{Rapid solution of problems by quantum
  computation,} {\protect\JournalTitle{Proceedings of the Royal Society of
  London. Series A: Mathematical and Physical Sciences}} \textbf{439}, 553--558
  (1992).

\bibitem{Simon97}
D.~R. Simon, \enquote{On the power of quantum computation,}
  {\protect\JournalTitle{SIAM journal on computing}} \textbf{26}, 1474--1483
  (1997).

\bibitem{Berstein97}
E.~Bernstein and U.~Vazirani, \enquote{Quantum complexity theory,}
  {\protect\JournalTitle{SIAM Journal on Computing}} \textbf{26}, 1411--1473
  (1997).

\bibitem{Shor97}
P.~Shor, \enquote{Polynomial-time algorithms for prime factorization and
  discrete logarithms on a quantum computer,} {\protect\JournalTitle{SIAM
  Journal on Computing}} \textbf{26}, 1484--1509 (1997).

\bibitem{Grover96}
L.~K. Grover, \enquote{Quantum mechanics helps in searching for a needle in a
  haystack,} {\protect\JournalTitle{Phys. Rev. Lett.}} \textbf{79}, 325--328
  (1997).

\bibitem{wan2017quantum}
K.~H. Wan, O.~Dahlsten, H.~Kristj{\'a}nsson, R.~Gardner, and M.~Kim,
  \enquote{Quantum generalisation of feedforward neural networks,}
  {\protect\JournalTitle{npj Quantum information}} \textbf{3}, 1--8 (2017).

\bibitem{farhi2018classification}
E.~Farhi and H.~Neven, \enquote{Classification with quantum neural networks on
  near term processors,} {\protect\JournalTitle{arXiv preprint
  arXiv:1802.06002}}  (2018).

\bibitem{rebentrost2018quantum}
P.~Rebentrost, T.~R. Bromley, C.~Weedbrook, and S.~Lloyd, \enquote{Quantum
  hopfield neural network,} {\protect\JournalTitle{Physical Review A}}
  \textbf{98}, 042308 (2018).

\bibitem{schuld2019quantum}
M.~Schuld and N.~Killoran, \enquote{Quantum machine learning in feature hilbert
  spaces,} {\protect\JournalTitle{Physical Review Letters}} \textbf{122},
  040504 (2019).

\bibitem{havlivcek2019supervised}
V.~Havl{\'\i}{\v{c}}ek, A.~D. C{\'o}rcoles, K.~Temme, A.~W. Harrow, A.~Kandala,
  J.~M. Chow, and J.~M. Gambetta, \enquote{Supervised learning with
  quantum-enhanced feature spaces,} {\protect\JournalTitle{Nature}}
  \textbf{567}, 209--212 (2019).

\bibitem{Horodecki09}
R.~Horodecki, P.~Horodecki, M.~Horodecki, and K.~Horodecki, \enquote{Quantum
  entanglement,} {\protect\JournalTitle{Rev. Mod. Phys.}} \textbf{81}, 865--942
  (2009).

\bibitem{Streltsov17}
A.~Streltsov, G.~Adesso, and M.~B. Plenio, \enquote{Colloquium: Quantum
  coherence as a resource,} {\protect\JournalTitle{Rev. Mod. Phys.}}
  \textbf{89}, 041003 (2017).

\bibitem{Bera17}
A.~Bera, T.~Das, D.~Sadhukhan, S.~S. Roy, A.~Sen(De), and U.~Sen,
  \enquote{Quantum discord and its allies: a review of recent progress,}
  {\protect\JournalTitle{Reports on Progress in Physics}} \textbf{81}, 024001
  (2017).

\bibitem{amaral2019resource}
B.~Amaral, \enquote{Resource theory of contextuality,}
  {\protect\JournalTitle{Philosophical Transactions of the Royal Society A}}
  \textbf{377}, 20190010 (2019).

\bibitem{Ferrie14}
C.~Ferrie and J.~Combes, \enquote{How the result of a single coin toss can turn
  out to be 100 heads,} {\protect\JournalTitle{Phys. Rev. Lett.}} \textbf{113},
  120404 (2014).

\bibitem{Allen1999iv}
L.~Allen, M.~Padgett, and M.~Babiker, \enquote{{IV} the orbital angular
  momentum of light,} {\protect\JournalTitle{Progress in optics}} \textbf{39},
  291--372 (1999).

\bibitem{andrews2012angular}
D.~L. Andrews and M.~Babiker, \emph{The angular momentum of light} (Cambridge
  University Press, 2012).

\bibitem{allen2011orbital}
L.~Allen and M.~Padgett, \enquote{The orbital angular momentum of light: An
  introduction,} {\protect\JournalTitle{Twisted Photons: Applications of Light
  with Orbital Angular Momentum}} pp. 1--12 (2011).

\bibitem{marrucci2011spin}
L.~Marrucci, E.~Karimi, S.~Slussarenko, B.~Piccirillo, E.~Santamato, E.~Nagali,
  and F.~Sciarrino, \enquote{Spin-to-orbital conversion of the angular momentum
  of light and its classical and quantum applications,}
  {\protect\JournalTitle{Journal of Optics}} \textbf{13}, 064001 (2011).

\bibitem{shen19}
Y.~Shen, X.~Wang, Z.~Xie, C.~Min, X.~Fu, Q.~Liu, M.~Gong, and X.~Yuan,
  \enquote{Optical vortices 30 years on: Oam manipulation from topological
  charge to multiple singularities,} {\protect\JournalTitle{Light: Science \&
  Applications}} \textbf{8}, 1--29 (2019).

\bibitem{tuan2018realizing}
P.-H. Tuan, H.-C. Liang, K.-F. Huang, and Y.-F. Chen, \enquote{Realizing
  high-pulse-energy large-angular-momentum beams by astigmatic transformation
  of geometric modes in an nd: Yag/{C}r4+: Yag laser,}
  {\protect\JournalTitle{IEEE Journal of Selected Topics in Quantum
  Electronics}} \textbf{24}, 1--9 (2018).

\bibitem{Mandel95}
L.~Mandel and E.~Wolf, \emph{Optical coherence and quantum optics} (Cambridge
  university press, 1995).

\bibitem{sztul2006double}
H.~Sztul and R.~Alfano, \enquote{Double-slit interference with
  laguerre-gaussian beams,} {\protect\JournalTitle{Optics Letters}}
  \textbf{31}, 999--1001 (2006).

\bibitem{Wolf07}
E.~Wolf \emph{et~al.}, \emph{Introduction to the Theory of Coherence and
  Polarization of Light} (Cambridge University Press, 2007).

\bibitem{jha2010angular}
A.~K. Jha, J.~Leach, B.~Jack, S.~Franke-Arnold, S.~M. Barnett, R.~W. Boyd, and
  M.~J. Padgett, \enquote{Angular two-photon interference and angular two-qubit
  states,} {\protect\JournalTitle{Physical Review Letters}} \textbf{104},
  010501 (2010).

\bibitem{Malik12measurement}
M.~Malik, S.~Murugkar, J.~Leach, and R.~W. Boyd, \enquote{Measurement of the
  orbital-angular-momentum spectrum of fields with partial angular coherence
  using double-angular-slit interference,} {\protect\JournalTitle{Physical
  Review A}} \textbf{86}, 063806 (2012).

\bibitem{Allen92}
L.~Allen, M.~W. Beijersbergen, R.~J.~C. Spreeuw, and J.~P. Woerdman,
  \enquote{Orbital angular momentum of light and the transformation of
  laguerre-gaussian laser modes,} {\protect\JournalTitle{Phys. Rev. A}}
  \textbf{45}, 8185--8189 (1992).

\bibitem{Forbes19}
A.~Forbes, A.~Aiello, and B.~Ndagano, \enquote{Classically entangled light,}
  {\protect\JournalTitle{Progress in Optics}} \textbf{64}, 99--153 (2019).

\bibitem{Stoklasa15}
B.~Stoklasa, L.~Motka, J.~Rehacek, Z.~Hradil, L.~L. S{\'a}nchez-Soto, and
  G.~Agarwal, \enquote{Experimental violation of a bell-like inequality with
  optical vortex beams,} {\protect\JournalTitle{New Journal of Physics}}
  \textbf{17}, 113046 (2015).

\bibitem{Borges10}
C.~V.~S. Borges, M.~Hor-Meyll, J.~A.~O. Huguenin, and A.~Z. Khoury,
  \enquote{Bell-like inequality for the spin-orbit separability of a laser
  beam,} {\protect\JournalTitle{Phys. Rev. A}} \textbf{82}, 033833 (2010).

\bibitem{Kagalwala13}
K.~H. Kagalwala, G.~Di~Giuseppe, A.~F. Abouraddy, and B.~E. Saleh,
  \enquote{Bell's measure in classical optical coherence,}
  {\protect\JournalTitle{Nature Photonics}} \textbf{7}, 72--78 (2013).

\bibitem{Hashemi15}
S.~M. Hashemi~Rafsanjani, M.~Mirhosseini, O.~S. Maga\~na Loaiza, and R.~W.
  Boyd, \enquote{State transfer based on classical nonseparability,}
  {\protect\JournalTitle{Phys. Rev. A}} \textbf{92}, 023827 (2015).

\bibitem{Li18}
T.~Li, X.~Zhang, Q.~Zeng, B.~Wang, and X.~Zhang, \enquote{Experimental
  simulation of monogamy relation between contextuality and nonlocality in
  classical light,} {\protect\JournalTitle{Opt. Express}} \textbf{26},
  11959--11975 (2018).

\bibitem{Goyal13}
S.~K. Goyal, F.~S. Roux, A.~Forbes, and T.~Konrad, \enquote{Implementing
  quantum walks using orbital angular momentum of classical light,}
  {\protect\JournalTitle{Phys. Rev. Lett.}} \textbf{110}, 263602 (2013).

\bibitem{eisaman2011invited}
M.~D. Eisaman, J.~Fan, A.~Migdall, and S.~V. Polyakov, \enquote{Invited review
  article: Single-photon sources and detectors,} {\protect\JournalTitle{Review
  of Scientific Instruments}} \textbf{82}, 071101 (2011).

\bibitem{Hemmer06}
P.~R. Hemmer, A.~Muthukrishnan, M.~O. Scully, and M.~S. Zubairy,
  \enquote{Quantum lithography with classical light,}
  {\protect\JournalTitle{Phys. Rev. Lett.}} \textbf{96}, 163603 (2006).

\bibitem{Kaur07}
G.~Kaur, G.~Narang \emph{et~al.}, \enquote{Optical implementations, oracle
  equivalence, and the bernstein-vazirani algorithm,}
  {\protect\JournalTitle{JOSA B}} \textbf{24}, 221--225 (2007).

\bibitem{Dixon09}
P.~B. Dixon, D.~J. Starling, A.~N. Jordan, and J.~C. Howell,
  \enquote{Ultrasensitive beam deflection measurement via interferometric weak
  value amplification,} {\protect\JournalTitle{Phys. Rev. Lett.}} \textbf{102},
  173601 (2009).

\bibitem{Atherton15}
D.~P. Atherton, G.~Ranjit, A.~A. Geraci, and J.~D. Weinstein,
  \enquote{Observation of a classical cheshire cat in an optical
  interferometer,} {\protect\JournalTitle{Opt. Lett.}} \textbf{40}, 879--881
  (2015).

\bibitem{Perezgarcia16}
B.~Perez-Garcia, M.~McLaren, S.~K. Goyal, R.~I. Hernandez-Aranda, A.~Forbes,
  and T.~Konrad, \enquote{Quantum computation with classical light:
  Implementation of the deutsch–jozsa algorithm,}
  {\protect\JournalTitle{Physics Letters A}} \textbf{380}, 1925 -- 1931 (2016).

\bibitem{Garcia18}
B.~Perez-Garcia, R.~I. Hernandez-Aranda, A.~Forbes, and T.~Konrad, \enquote{The
  first iteration of grover's algorithm using classical light with orbital
  angular momentum,} {\protect\JournalTitle{Journal of Modern Optics}}
  \textbf{65}, 1942--1948 (2018).

\bibitem{Konrad19}
T.~Konrad and A.~Forbes, \enquote{Quantum mechanics and classical light,}
  {\protect\JournalTitle{Contemporary Physics}} \textbf{60}, 1--22 (2019).

\bibitem{Zhang19}
S.~Zhang, P.~Li, B.~Wang, Q.~Zeng, and X.~Zhang, \enquote{Implementation of
  quantum permutation algorithm with classical light,}
  {\protect\JournalTitle{Journal of Physics Communications}} \textbf{3}, 015008
  (2019).

\bibitem{Chevalier21}
H.~Chevalier, A.~J. Paige, H.~Kwon, and M.~S. Kim, \enquote{Violating the
  leggett-garg inequalities with classical light,} {\protect\JournalTitle{Phys.
  Rev. A}} \textbf{103}, 043707 (2021).

\bibitem{Bharath14}
H.~M. Bharath and V.~Ravishankar, \enquote{Classical simulation of entangled
  states,} {\protect\JournalTitle{Phys. Rev. A}} \textbf{89}, 062110 (2014).

\bibitem{Asthana21}
S.~Asthana, R.~Bala, and V.~Ravishankar, \enquote{Quantum communication with
  {SU (2)} invariant separable $2\times$ {N} level systems,}
  {\protect\JournalTitle{arXiv preprint arXiv:2104.04469}}  (2021).

\bibitem{Adhikary20a}
S.~Adhikary, S.~Dangwal, and D.~Bhowmik, \enquote{Supervised learning with a
  quantum classifier using multi-level systems,} {\protect\JournalTitle{Quantum
  Information Processing}} \textbf{19}, 1--12 (2020).

\bibitem{Zhang2020review}
K.~Zhang, Y.~Wang, Y.~Yuan, and S.~N. Burokur, \enquote{A review of orbital
  angular momentum vortex beams generation: from traditional methods to
  metasurfaces,} {\protect\JournalTitle{Applied Sciences}} \textbf{10}, 1015
  (2020).

\bibitem{Wang17}
Y.~Wang, V.~Poto{\v{c}}ek, S.~M. Barnett, and X.~Feng, \enquote{Programmable
  holographic technique for implementing unitary and nonunitary
  transformations,} {\protect\JournalTitle{Physical Review A}} \textbf{95},
  033827 (2017).

\bibitem{Spreeuw01}
R.~J.~C. Spreeuw, \enquote{Classical wave-optics analogy of quantum-information
  processing,} {\protect\JournalTitle{Phys. Rev. A}} \textbf{63}, 062302
  (2001).

\bibitem{landau2013classical}
L.~D. Landau, \emph{The classical theory of fields}, vol. 2, pp 119-124
  (Elsevier, 2013).

\bibitem{Adhikary16}
S.~Adhikary, I.~K. Panda, and V.~Ravishankar, \enquote{Super-quantum states in
  su(2) invariant level systems,} {\protect\JournalTitle{Annals of Physics}}
  \textbf{377}, 85--95 (2017).

\bibitem{Radcliffe71}
J.~M. Radcliffe, \enquote{Some properties of coherent spin states,}
  {\protect\JournalTitle{Journal of Physics A: General Physics}} \textbf{4},
  313--323 (1971).

\bibitem{Guzman16}
D.~Guzman-Silva, R.~Br{\"u}ning, F.~Zimmermann, C.~Vetter, M.~Gr{\"a}fe,
  M.~Heinrich, S.~Nolte, M.~Duparr{\'e}, A.~Aiello, M.~Ornigotti \emph{et~al.},
  \enquote{Demonstration of local teleportation using classical entanglement,}
  {\protect\JournalTitle{Laser \& Photonics Reviews}} \textbf{10}, 317--321
  (2016).

\bibitem{Milione12}
G.~Milione, S.~Evans, D.~A. Nolan, and R.~R. Alfano, \enquote{Higher order
  pancharatnam-berry phase and the angular momentum of light,}
  {\protect\JournalTitle{Phys. Rev. Lett.}} \textbf{108}, 190401 (2012).

\bibitem{Aiello15}
A.~Aiello, F.~T{\"o}ppel, C.~Marquardt, E.~Giacobino, and G.~Leuchs,
  \enquote{Quantum- like nonseparable structures in optical beams,}
  {\protect\JournalTitle{New Journal of Physics}} \textbf{17}, 043024 (2015).

\bibitem{Chen21}
L.~Chen, \enquote{Quantum discord of thermal two-photon orbital angular
  momentum state: mimicking teleportation to transmit an image,}
  {\protect\JournalTitle{Light: Science \& Applications}} \textbf{10}, 1--8
  (2021).

\bibitem{Balthazar21}
W.~F. Balthazar, D.~G. Braga, V.~S. Lamego, M.~H.~M. Passos, and J.~A.~O.
  Huguenin, \enquote{Spin-orbit $x$ states,} {\protect\JournalTitle{Phys. Rev.
  A}} \textbf{103}, 022411 (2021).

\bibitem{Byrd_1998}
M.~Byrd and E.~Sudarshan, \enquote{{SU} (3) revisited,}
  {\protect\JournalTitle{Journal of Physics A: Mathematical and General}}
  \textbf{31}, 9255 (1998).

\bibitem{Forbes16}
A.~Forbes, A.~Dudley, and M.~McLaren, \enquote{Creation and detection of
  optical modes with spatial light modulators,} {\protect\JournalTitle{Advances
  in Optics and Photonics}} \textbf{8}, 200--227 (2016).

\end{thebibliography}

\end{document}